\documentclass[manuscript]{aastex63}
\usepackage{amsmath}
\shorttitle{Helicity}
\shortauthors{Yang et al.}
\begin{document}
\title{Relative Magnetic Helicity  Based on a Periodic Potential Field}
\email{k.yang@sydney.edu.au}
\author[0000-0002-7663-7652]{Kai E.~Yang}
\author[0000-0001-5100-2354]{Michael~S.~Wheatland}
\affil{Sydney Institute for Astronomy, School of Physics, The University of Sydney, Sydney, NSW 2006, Australia}
\author{Stuart~A.~Gilchrist}
\affil{NorthWest Research Associates, Boulder, CO 80301, USA}

\begin{abstract}
Magnetic helicity is conserved under ideal magnetohydrodynamics (MHD) and quasi-conserved even under a resistive process.
The standard definition for magnetic helicity cannot be applied directly to an open magnetic field in a volume, because it is gauge-dependent.
Instead, the relative magnetic helicity is widely used.
We find that the energy of a potential magnetic field in a rectangular domain with periodic lateral boundary conditions is less than that of the field with a fixed normal component on all six boundaries.
To make use of this lower energy potential field in the analysis of relative magnetic helicity, we introducing a new definition for magnetic helicity for the magnetic field, which involves the periodic potential field.
We apply this definition to a sequence of analytic solutions and a numerical simulation.
The results show that our new gauge-invariant helicity is very close to the current-carrying part of the relative magnetic helicity of the original magnetic field.
We find also that the ratio between the current-carrying helicity and the relative magnetic helicity for the original and our defined relative helicity show different behavior.
It seems that the new helicity is more sensitive to the component of the field due to the electric current in the volume, which is the source for instabilities and solar eruptive phenomena.
\end{abstract}


\section{Introduction}\label{sec:1}
Magnetic helicity is a global measurement of the magnetic field line linkage in a closed volume where the normal component of the magnetic field vanishes on the boundary \citep{Woltjer1958a,Woltjer1958b,Moffatt1969}.
One of the most important properties of the magnetic helicity is that it is strictly invariant under an ideal magnetohydrodynamic (MHD) process, and quasi-invariant under resistive MHD \citep{Berger1984b,Taylor1986,Berger1995}.
However, in the case where magnetic field lines cross the boundary, like the case of the solar atmosphere, the magnetic helicity is not gauge-independent.
This issue was solved by introducing a new definition of helicity, the relative magnetic helicity, which uses the difference between the real magnetic field in the unbounded volume and a reference field defined in the same volume that shares the same normal component of the magnetic field on all boundaries \citep{Berger1984a}.

A potential field is usually chosen as the reference field.
The potential magnetic field ${\bf B}_{\rm p}$ is based on the hypothesis that there is no electric current in the volume, i.e.~$\nabla \times {\bf B}_{\rm p}=0$.
Thus the magnetic field can be written as the gradient of a scalar field, ${\bf B}_{\rm p} = \nabla \phi$, where $\phi$ satisfies Laplace's equation, $\nabla^2 \phi = 0$.
If boundary conditions are available on all boundaries, then the Neumann boundary condition can be used, i.e.~$\partial_{\rm n}\phi|_{\partial\Omega}=B_{\rm n}|_{\partial\Omega}$ on all boundaries, where $\Omega$ and $\partial\Omega$ denote the computational domain and its boundary, and $B_n$ is the normal component of the field.
This choice of a potential field for the reference field is required for the relative magnetic helicity to be gauge invariant.
According to Thomson's theorem \citep{Stratton1941,Sakurai1979}, the potential field with the Neumann boundary condition on all boundaries is the minimum energy field, for those boundary conditions.

However, in the case of magnetic fields on the Sun, only the photospheric and/or chromospheric magnetograms can be obtained from observations and serve as the bottom boundary.
When potential fields are calculated from solar boundary conditions, some assumptions should be made for the lateral and top boundaries, e.g.~a periodic condition on the lateral boundaries.
Usually, the potential magnetic field ${\bf B}_{\rm p}$ can be obtained by using the Green's function technique \citep{Chiu1977} or Fourier transformation \citep{Alissandrakis1981}.
The Fourier technique intrinsically involves periodic boundary conditions.

When calculating relative magnetic helicity, we should keep in mind that the potential field is not the only choice of the reference field: any field shares the same normal component of the magnetic field on the boundaries can play the role of the reference field.
Various authors have considered the properties and definition of relative magnetic helicity.
\cite{2014Prior} proved the existence of an untwisted reference field.
\cite{Low2006} proposed a primitive form of the magnetic helicity based on the Chandrasekhar-Kendal decomposition of the magnetic field.
In general the relative magnetic helicity is only uniquely defined if we restrict the choice of the reference field.
It worth noting that solving the Laplace's equation with Neumann boundary conditions with an irregular boundary and/or a non-uniform grid is complex and challenging, and various techniques have been proposed \citep{2008Longcope,2009Malanushenko,Teunissen2019}.
Some additional properties of relative magnetic helicity might exist depending on the choice of the reference field, e.g.~the conservation property of the helicity \citep{Pariat2015}.
However, this topic is outside the scope of this paper.

The periodic potential magnetic field ${\bf B}_0$ used in the force-free extrapolation methods, like the Current-field Iteration Method \citep{Wheatland2006,Wheatland2007}, has a lower energy than the potential field with Neumann boundary condition on all boundaries if the original magnetic field is itself periodic, and has equal net fluxes on the top and bottom boundaries, as demonstrated in Appendix \ref{sec:a:1}.
The definition and uniqueness of the periodic potential field are demonstrated in Appendix \ref{sec:a:2}.
Hence it is of interest to consider the use of this field in defining relative magnetic helicity.

A given magnetic field ${\bf B}$ can be decomposed as the sum of a potential field ${\bf B}_{\rm p}$ and a current-associated field ${\bf B}_{\rm j}$, where ${\bf B}_{\rm p}$ comes from Laplace's equation with Neumann boundary conditions, and ${\bf B}_{\rm j}$ is the residual field.
If the vector potential of the magnetic field experiences a gauge transform, ${\bf A}\rightarrow {\bf A}+\nabla\psi$, the change in the relative magnetic helicity is $\Delta H_{\rm r} = \int  \psi({\bf B}-{\bf B}_{\rm p}) \cdot~\rm{d}{\bf S}$.
Using the Neumann boundary condition of ${\bf B}_{\rm p}$, the surface integral is zero, which ensures the relative magnetic helicity is gauge-invariant.
However, the periodic potential field ${\bf B}_0$ only depends on the top and bottom data and assumes the lateral boundary is periodic.
Hence the current-associated magnetic field is not closed.
As a consequence of this, $\Delta H_{\rm r}$ will not be zero, breaking the gauge-invariant property.
This demonstrates that the periodic potential field cannot play the role of the reference field directly for the relative magnetic helicity, with the usual definition.
 
In this paper, we present a new definition for a relative magnetic helicity partly based on the periodic potential field.
The newly defined magnetic helicity is consistent with the result from \cite{Berger1997}, in that with the newly defined helicity the periodic potential field is not used as the reference field directly.

This paper is organized as follows. 
Section \ref{sec:2} introduces the new definition, and we apply the new concept to both static and dynamic magnetic models in Section \ref{sec:3}. In Section \ref{sec:4}, we summarize results on the new magnetic helicity based on the periodic potential field.

\section{definition of helicity based on periodic potential field}\label{sec:2}

For comparison with our new definition, we will briefly review the relative magnetic helicity given by \cite{Berger1984a}.
A magnetic field ${\bf B}$ in a three-dimension (3D) volume, $\Omega$, can be decomposed as ${\bf B}={\bf B}_{\rm j}+{\bf B}_{\rm p}$, with the boundary condition $({\bf B}-{\bf B_{\rm p}})\cdot\hat{\bf n}|_{\partial \Omega}=0$, where $\partial\Omega$ is the boundary and $\hat{\bf n}$ is the associated unit normal vector.
Thus ${\bf B}_{\rm p}$ can play the role of the reference field.
With this decomposition, the relative magnetic helicity can be defined with the formula from \cite{Finn1985},
\begin{equation}\label{eq:1}
H_{\rm r}=\int_{\Omega}({\bf A}+{\bf A}_{\rm p})\cdot({\bf B}-{\bf B}_{\rm p}){\rm d^3}{\bf x},
\end{equation}
where ${\bf A}$ and ${\bf A}_{\rm p}$ are the corresponding vector potentials.
The above formula is widely used in both theoretical and numerical computation \citep{Demoulin2003,Demoulin2006,2008Longcope,Jing2012,Yang2013,Yang2018,Pariat2005,Pariat2015,Pariat2017,Valori2012,Valori2016,Guo2017,Moraitis2019}.
\cite{Berger1999} separated the relative magnetic helicity of Equation (\ref{eq:1}) into two gauge independent parts, the current-carrying part, $H_{\rm j}$, and the mutual helicity between the potential and current-carrying fields, $H_{\rm pj}$.
Specifically, $H_{\rm r} = H_{\rm j} + H_{\rm pj}$, with
\begin{equation}\label{eq:2}
H_{\rm j} =\int_{\Omega}({\bf A}-{\bf A}_{\rm p})\cdot({\bf B}-{\bf B}_{\rm p}){\rm d^3}{\bf x},
\end{equation}
and
\begin{equation}\label{eq:3}
H_{\rm pj}=2\int_{\Omega} {\bf A}_{\rm p}\cdot({\bf B}-{\bf B}_{\rm p}){\rm d^3}{\bf x}.
\end{equation}
Recently, based on the analysis of numerical simulations and observations of eruptions in the solar corona, it has been hypothesized that the ratio between $H_{\rm j}$ and $H_{\rm r}$ might have a crucial value for the onset of an eruption \citep{Pariat2017,Zuccarello2018,Linan2018,Moraitis2019}.

We can also decompose a 3D magnetic field ${\bf B}$ into a current-associated field and a potential field with periodic boundary condition, ${\bf B}={\bf B}_0 + {\bf B}_{\rm c}$, following the procedure in the Current-field Iteration Method for extrapolation of non-linear force-free fields from bottom boundary data \citep{Wheatland2006,Wheatland2007}.
In this case, ${\bf B}_0$ satisfies the condition $({\bf B}_0 - {\bf B})\cdot {\bf \hat{n}}=0$ on the bottom and top boundaries, and we assume all of the lateral boundaries are periodic.
Therefore ${\bf B}_0$ does not match the lateral boundary condition on ${\bf B}$, and ${\bf B}_{\rm c}\cdot\hat{\bf n}$ on the lateral boundaries does not vanish.
The field ${\bf B}_0$ is uniquely defined ignoring the possibility of a constant horizontal field, as demonstrated in Appendix \ref{sec:a:2}.
The possibility of a constant horizontal field component is usually neglected during extrapolation, because the calculation of the periodic potential field only depends on the top and bottom boundaries.
As previously stated, it is not possible to use the periodic potential field, ${\bf B}_0$, as a reference field for calculating relative magnetic helicity, because the result is gauge dependent. However, following the original definition from \cite{Berger1984a}, we can decompose ${\bf B}_{\rm c}$ into two parts, ${\bf B}_{\rm c}={\bf B}_{\rm c1}+{\bf B}_{\rm p1}$, where ${\bf B}_{\rm p1}$ is the solution of Laplace's equation that satisfies the boundary condition $({\bf B}_{\rm c}-{\bf B}_{\rm p1})\cdot\hat{\bf n}|_{\partial \Omega}=0$ on all boundaries.
Similar to Equation (\ref{eq:1}) from \cite{Finn1985}, we can then define a gauge-invariant relative magnetic helicity for the field ${\bf B}_{\rm c}$:
\begin{equation}\label{eq:4}
H_{\rm cr} = \int_{\Omega}({\bf A}_{\rm c}+{\bf A}_{\rm p1})\cdot({\bf B}_{\rm c}-{\bf B}_{\rm p1}){\rm d^3}{\bf x}.
\end{equation}
Following the definition in Equations (\ref{eq:2}) and (\ref{eq:3}), we have $H_{\rm cr} = H_{\rm cj} + H_{\rm cpj}$, where
\begin{equation}\label{eq:5}
H_{\rm cj} = \int_{\Omega}({\bf A}_{\rm c}-{\bf A}_{\rm p1})\cdot({\bf B}_{\rm c}-{\bf B}_{\rm p1}){\rm d^3}{\bf x},
\end{equation}
and
\begin{equation}
H_{\rm cpj} = 2\int_{\Omega}{\bf A}_{\rm p1}\cdot({\bf B}_{\rm c}-{\bf B}_{\rm p1}){\rm d^3}{\bf x},
\end{equation}
where ${\bf A}_{\rm p1}$ and ${\bf A}_{\rm c}$ are the corresponding vector potentials for the magnetic fields ${\bf B}_{\rm p1}$ and ${\bf B}_{\rm c}$, respectively.
Evidently, both $H_{\rm cpj}$ and $H_{\rm cj}$ are gauge-invariant.

Both ${\bf B}_{\rm c1}$ and ${\bf B}_{\rm j}$ obey Ampere's Law:
\begin{equation}
\nabla\times{\bf B}_{\rm c1}=\nabla\times{\bf B}_{\rm j}=\mu_0{\bf J},
\end{equation}
where ${\bf J}$ is the current density in the volume.
From the boundary condition of ${\bf B}_{\rm c}$ and ${\bf B}_{\rm p1}$, we can find that the current-associated magnetic field ${\bf B}_{\rm c1}$ 
satisfies ${\bf B}_{\rm c1}\cdot\hat{\bf n}|_{\partial\Omega}=0$, which is the same boundary condition as ${\bf B}_{\rm j}$.
Because ${\bf B}_{\rm c1}$ and ${\bf B}_{\rm j}$ satisfy the same partial differential equation with the same boundary conditions, we must have ${\bf B}_{\rm c1} = {\bf B}_{\rm j}$.
Comparing the two decompositions, ${\bf B} = {\bf B}_{\rm p} +  {\bf B}_{\rm j}$ and ${\bf B} = {\bf B}_{\rm c} +  {\bf B}_{\rm 0} =  {\bf B}_{\rm c1} +  {\bf B}_{\rm p1} + {\bf B}_{\rm 0}$, two relations can be obtained: ${\bf B}_{\rm p} = {\bf B}_{\rm 0} +{\bf B}_{\rm p1}$ and ${\bf B}_{\rm c} = {\bf B}_{\rm j} +{\bf B}_{\rm p1}$.
Using Equations (\ref{eq:2}) and (\ref{eq:5}), it is easy to see that $H_{\rm cj}$ is exactly the same as $H_{\rm j}$.
From the definitions, Equations (\ref{eq:1}) and (\ref{eq:4}), we find that if ${\bf B}_{\rm c}$ vanishes on all boundaries, the potential field ${\bf B}_{\rm p1}$ will be zero, thus ${\bf B}_{\rm c}={\bf B}_{\rm j}$.
Then $H_{\rm cr}$  and $H_{\rm cj}$ reduce to $H_{\rm j}= \int{\bf A}_{\rm j}\cdot{\bf B}_{\rm j}~{\rm d^3}{\bf x}$ and $H_{\rm cpj} = 0$.
Strictly speaking, the field ${\bf B}_{\rm p1}$ is the potential field corresponding to the helicity $H_{\rm cr}$, rather than ${\bf B}_0$.

For calculating the helicity, we need to compute the vector potential corresponding to each part of the magnetic field.
The periodic potential and the current-associated fields can be calculated from a Fourier technique, and thus the associated vector potential can be easily computed \citep{Wheatland2007}.
We compute $\mathbf A_{\rm p1}$ in the Coulomb gauge by solving the vector Poisson equation numerically. 
Appendix \ref{sec_sag_appendix} describes our formulation of a boundary-value problem for $\mathbf A_{\rm p1}$ and its solution.

\section{Application and Comparison}\label{sec:3}

\subsection{Titov-D\'{e}moulin model}\label{sec:3:1}
\begin{figure}
\centering
\includegraphics[width=0.9\textwidth]{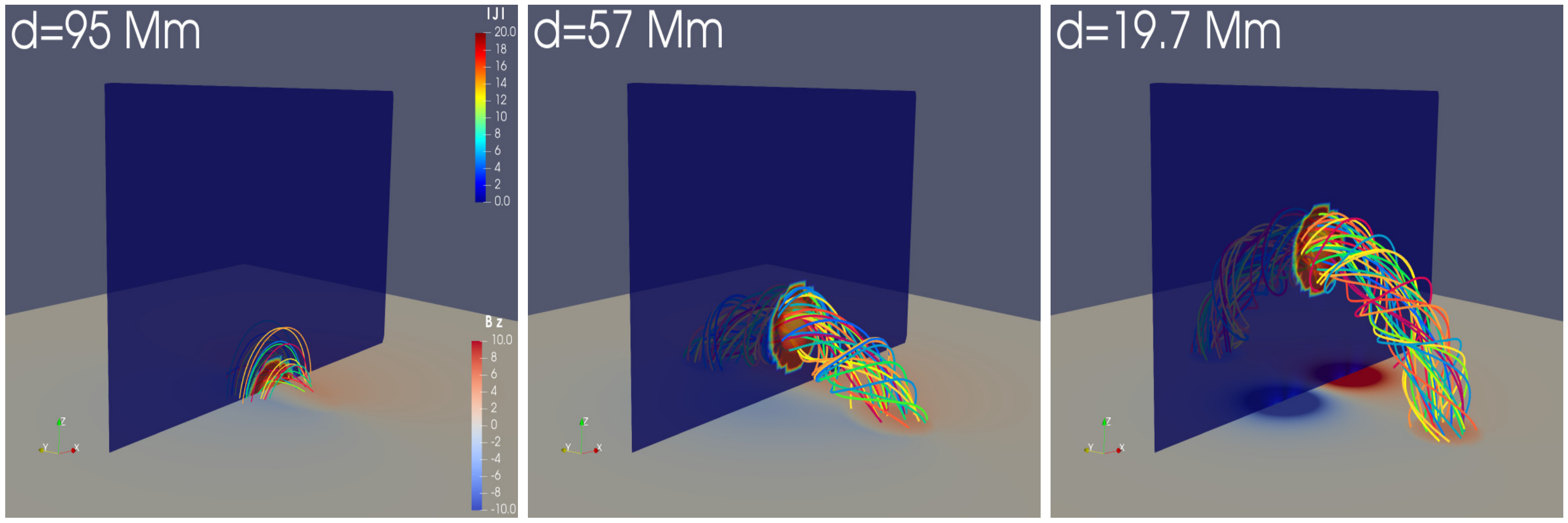}
\caption{Magnetic field for the artificial emergence process mimicked by changing the value of the parameter $d$ in a sequence of Titov-D\'{e}moulin equilibrium solutions.
The bottom boundary shows the distribution of $B_z$, and the vertical slice shows the total current density $|{\bf J}|$.
The values of the magnetic field and current are in units of $5.9$ gauss and $8.8\times10^{20}$ statampere.
The colored lines indicate magnetic field lines associated with the flux rope.}\label{fig:1}
\end{figure}

We test our new helicity on a series of Titov-D\'{e}moulin (TD) flux-rope models \citep{Titov1999}.
The parameters are $L=35$ Mm, $R=110$ Mm, $a=23.9$ Mm, $l_i=0.5$, $q=40$ T Mm$^2$, and a range of values of $d$ from $1$ to $135$ Mm.
The computational domain in the range of $-300~{\rm Mm}<x<300~{\rm Mm}$, $-300~{\rm Mm}<y<300~{\rm Mm}$, $0~{\rm Mm}<z<600~{\rm Mm}$, on a uniform grid with size  $128^3$, in this domain, the net flux is zero at both bottom and top boundaries.
As the parameter $d$ indicates the depth of the axis of the flux rope, the decrease of this parameter mimics an artificial emergence process for the current system.
With the pseudo-emergence, helicity and energy are injected into the computational domain similar to what happens during the emergence of a solar active region \citep{Liu2014}.
When the parameter $d$ is less than 70 Mm, the flux rope is unstable, but since this is not our main topic, we will not further discuss the stability analysis.
However, there is no physical flow on the boundary, so the associated injection flux cannot be calculated directly.
Three snapshots of the flux rope emergence process are shown in Figure \ref{fig:1}.
With the rise of the magnetic flux rope, the background magnetic field also increases simultaneous with the decrease of the parameter $d$.

We show the spatial integral of energy and helicity with the decrease of parameter $d$ in Figure \ref{fig:2}.
The magnetic energy and helicity are analysed using our decompositions based on the periodic potential field ${\bf B}_0$ (blue lines), and the potential field with its normal component fixed on six boundaries ${\bf B}_{\rm p}$ (red lines).
We show the total magnetic energy and the current-carrying part of the relative magnetic helicity with black solid lines. 
From the evolution of the energy (Figure \ref{fig:1} (a)), the main feature is that all components of the energy increase, which is expected due to the current system emerging into the computational domain.
We find that the energy from the periodic potential field, $E_{0}=\frac{1}{8\pi}\int_{\Omega}B^2_0~{\rm d}^3 {\bf x}$, is slightly less than that of the potential field ${\bf B}_{\rm p}$, $E_{\rm p}=\frac{1}{8\pi}\int_{\Omega}B^2_{\rm p}~{\rm d}^3 {\bf x}$, as expected from the Thomson theorem (see Appendix \ref{sec:a:1}).
It worth mentioning that in Appendix \ref{sec:a:1}, the proof assumes that the original magnetic field is also periodic.
However, from the test in Appendix \ref{sec:a:3}, even when the magnetic field $\bf B$ does not have a periodic lateral boundary condition, $E_0$ can still be smaller than $E_{\rm p}$.
The energies of the two potential fields are very close to each other and co-evolve during the artificial emergence.
When the parameter $d$ further decreases, each part of the energy increases dramatically, in particular, the potential energy, since the magnetic charges and the line current along the flux rope axis in the TD model are then close to the bottom boundary.

On the other hand, the helicity shows a rather different behavior than energy.
The helicity $H_{\rm cpj}$ is much smaller than $H_{\rm pj}$.
Thus the relative magnetic helicity based on the periodic potential field, $H_{\rm cr}$ is very close to the current-carrying part of the magnetic helicity $H_{\rm j}$.
This is a feature of the newly defined helicity.
Comparing the new helicity with the original one, we can find that the value of $H_{\rm pj}$ departs from $H_{\rm r}$, and $H_{\rm j}$ gradually dominates the relative magnetic helicity.
The new relative helicity $H_{\rm cr}$ always follows the mutual helicity between the current-carrying field and the potential field, $H_{\rm cpj}$.

\subsection{Eruptive case}\label{sec:3:2}
Magnetic helicity plays an important role in solar eruptions \citep{Low1996}.
This has been shown in many previous studies using both observation and theory \citep{Park2008,Park2010,Pariat2017,Zuccarello2018,Linan2018,Moraitis2019}.
We have applied both our newly defined and the original relative magnetic helicity to the data from an eruption process model, an isothermal MHD simulation with same settings as in \cite{Mei2017}, which uses the Message Passing Interface Adaptive Mesh Refinement Versatile Advection Code \citep{Keppens2003,Keppens2012,Porth2014,Xia2018}.
The initial condition of this MHD simulation is the TD model with the same parameters as in Section \ref{sec:3:1} but with the constant value $d=30$ Mm.
For the boundary conditions, the magnetic field is fixed at ghost layers at the initial value, and the velocity in the ghost layer of the bottom boundary is determined by a constant value extrapolation whilst the velocity at the other boundaries is fixed at zero.
Due to the symmetry of the magnetic field, the net flux remains zero at both top and bottom boundaries during the simulation.

We show three snapshots of the magnetic field during the eruptive process in Figure \ref{fig:3}.
As the twist of this initial condition exceeds the Kruskal-Shafranov condition, the kink instability sets in immediately at the start of the simulation.
A current sheet forms with the rise of the magnetic flux rope, however, since this is not our main focus we use a coarser mesh than that used in the original simulation \citep{Mei2017}.
The computational domain and resolution are the same as that used in Section \ref{sec:3:1}.

The total energy and helicity in the volume is shown in Figure \ref{fig:4}.
As the bottom flow is not zero, both a Poynting flux and helicity flux can be injected into the computational domain, so the absolute value of both energy and helicity increase with the development of the whole eruption (Figure \ref{fig:4}), similar to the increase seen in the artificial emergence in Section \ref{sec:3:1}.
The normal component of the magnetic field on the lower boundary does not change during the simulation.
Hence the periodic potential field is almost unchanged, since it only depends on the distribution of $B_z$ on the bottom and top boundaries, and the top boundary is so high that the magnetic field on it is very small.
The relation with ${\bf B}_0$ is shown in Appendix B.
As a consequence of this, the energy of the periodic potential field, $E_0$, is constant (Figure \ref{fig:4} (a)).
The normal component of the magnetic field on the lateral boundaries changes a little, which leads to $E_{\rm p}$ becoming a little larger than $E_0$.
Regarding the magnetic helicity, the potential field component, ${\bf B}_{\rm p1}$ is very small, which makes $H_{\rm cpj}$ close to zero because it is the coupling between this component of the potential field and the current-carrying part.
Therefore, the value of $H_{\rm cr}$ is very close to that of $H_{\rm j}$ and shows a monotonic increase during the whole simulation period.
The helicity associated with the potential field ${\bf B}_{\rm p}$ does not show a departure from $H_{\rm r}$ in this case, in contrast to the artificial emergence process in Section \ref{sec:3:1}.
That is due to the boundary conditions: $B_z$ is fixed in this case; whereas in the emergence case, $B_z$ changes due to the line current and magnetic charges approaching the lower boundary.

\section{summary and discussion}\label{sec:4}
In this paper, we proposed a relative magnetic helicity based on a periodic potential field, but we do not use the potential field directly as the reference field.
Our new helicity has a close relationship with the original formula from \cite{Finn1985}.
We should mention that from the definition, this new magnetic helicity can only be applied to the case where the domain has a periodic lateral boundary, e.g.~a Cartesian box.
It does not apply, e.g.~to a cylindrical domain.
Moreover, equal magnetic net fluxes on the top and bottom boundaries are required.
We apply the new helicity to two cases.
One is a series of calculations of the TD model, which mimic an artificial emergence of a magnetic flux rope.
The other is a dataset from an isothermal MHD simulation for a magnetic flux rope eruption.
The absolute value of energy and helicity in both cases show an increase during the development of the current system (Figure \ref{fig:2} and \ref{fig:4}). 
The most important difference between the original helicity and our definition is that our one is much closer to the mutual helicity between the current-carrying part of the field and the potential field, $H_{\rm pj}$.
We also make a further check in Appendix \ref{sec:a:3} on the calculation by using the magnetic field in half of the computational domain by making a slice through the flux rope.
The result (Figure \ref{fig:7}) is similar to that obtained in Section \ref{sec:3}.
It is worthwhile to mention that the lower energy state of the periodic potential field derived in Appendix \ref{sec:a:1} is based on a periodic current-associated magnetic field ${\bf B}_{\rm c}$.
This is not the case in TD model, however, it is still true in the calculation of the test cases, especially, the case in Appendix \ref{sec:a:3}, which cuts the domain into two parts so that the magnetic flux rope crosses the boundary and the magnetic field on the boundary is not small.
The relative energy difference between $E_0$ and $E_{\rm p}$ is much larger than the error level in \cite{Valori2013}.
Besides, there is another freedom of the proposed relative magnetic helicity, the choice of the top and bottom boundaries.
Obviously, in our test case, we use a natural choice of the bottom and top boundaries of the TD model.
Consider rotating by $90^{\circ}$ into a new coordinate system $x'=x$, $y'=-z$, and $z'=y$.
The normal component of the magnetic field on the new bottom and top boundaries will be very small, since the computational domain is very large, and hence the new ${\bf B}_0'$ will be very small.
As a consequence of this, $H_{\rm cr}'$ and $H_{\rm cpj}'$ will be very close to $H_{\rm r}'$ and $H_{\rm pj}'$, respectively.
All the variables with a prime indicate the corresponding variables in the rotated coordinate system.

\begin{figure}
\centering
\includegraphics[width=1\textwidth]{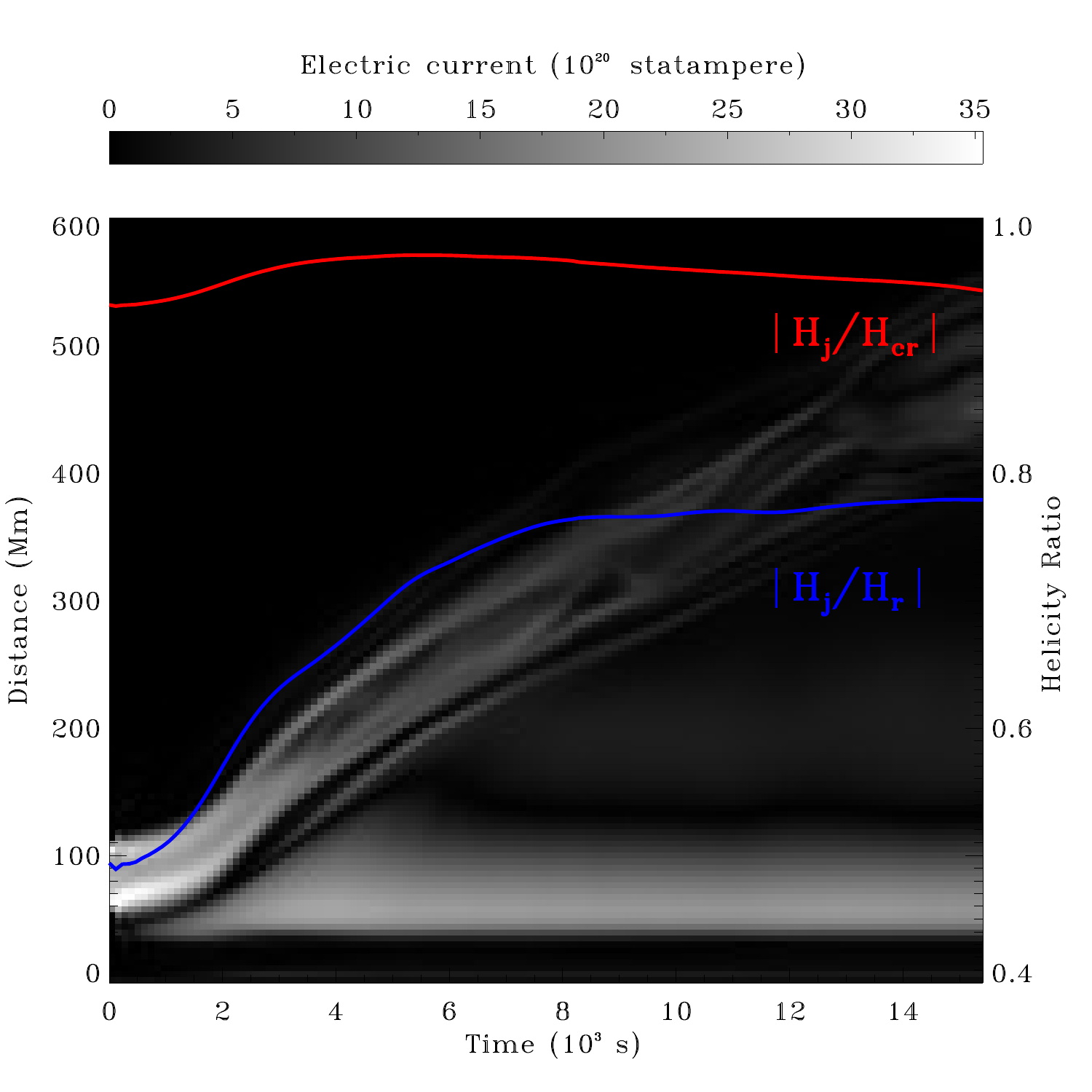}
\caption{The background shows the time-distance diagram of the electric current from the eruption simulation from Section \ref{sec:3:2} along the line from the bottom to the top of the simulation, at the center of the $\textit{x--y}$ plane.
The red and blue curves indicate the helicity ratio in our new definition, $|H_{\rm j}/H_{\rm cr}|$, and the original one, $|H_{\rm j}/H_{\rm r}|$, respectively.}\label{fig:5}
\end{figure}

In recent research, it has been argued that the ratio between the current-carrying helicity and the relative helicity might play a crucial role for the onset of a solar eruption \citep{Pariat2017,Zuccarello2018,Linan2018,Moraitis2019}.
Simulations suggest that this ratio increases just before solar flares and relaxes after \citep{Pariat2017,Moraitis2019}.
Figure \ref{fig:5} shows this ratio for the original relative magnetic helicity and also the ratio with our definition for the eruptive case.
The background of Figure \ref{fig:5} is the time-distance diagram of the electric current along the line from the bottom to the top at the center of the \textit{x--y} plane of the computational center.
The ratio for the new definition $|H_{\rm j}/H_{\rm cr}|$ experiences a gradual increase followed by an almost constant stage, whereas for $|H_{\rm j}/H_{\rm r}|$, a peak appears around $5\times 10^3$ s with the magnetic flux rope rising.
Both curves increase at the initial stage when the current system is rising due to the kink instability.
Thus our newly calculated relative magnetic helicity offers a new tool to investigate the MHD system, that might be more closely related to the current.

We have presented a new definition for magnetic helicity, which shows different behaviour to the usual relative magnetic helicity in test cases.
This indicates that magnetic helicity for open magnetic fields is not a uniquely defined quantity.

To better understand the relative magnetic helicity and make comparison with previous researches \citep{2014Prior,2019Prior,Yang2013,Yang2018,Pariat2017,Zuccarello2018,Linan2018,Moraitis2019}, we need to make further detail analyses by using our definition on more general cases and comparing it with the original helicity.
Moreover, the physical role of helicity should be investigated in detail for the onset of an MHD instability and the following eruptive process \citep{Guo2017,Pariat2017,Zuccarello2018}.
It is also of interest to apply our new definition to the magnetic field reconstructed from observed magnetograms.

\acknowledgments
This work was funded in part by an Australian Research Council Discovery Project (DP180102408).
Kai E.~Yang thanks Dr.~Z.~X.~Mei, for the discussion on simulation.
S.~A.~Gilchrist acknowledges that this material is based upon work supported by the National Science Foundation under Grant No. 1841962. 
Any opinions, findings, and conclusions or recommendations expressed in this material are those of the authors and do not necessarily reflect the views of the National Science Foundation. 

\newpage
\appendix
\section{Energy of Periodic Potential Field}\label{sec:a:1}

The Thomson theorem involves the decomposition of a magnetic field in a volume $\Omega$ as 
\begin{equation}\label{aeq:1}
{\bf B} = {\bf B}_{\rm j} + {\bf B}_{\rm p},
\end{equation}
where ${\bf B}_{\rm p}=\nabla\phi$ is the potential field satisfying 
\begin{equation}\label{aeq:2}
{\bf B}_{\rm p}\cdot\hat{\bf n}|_{\partial\Omega}={\bf B}\cdot\hat{\bf n}|_{\partial\Omega}
\end{equation}
on all boundaries.
The energy of the field is
\begin{equation}\label{aeq:3}
\begin{aligned}
E&=\frac{1}{8\pi}\int_{\Omega}({\bf B}_{\rm j} + {\bf B}_{\rm p})\cdot({\bf B}_{\rm j} + {\bf B}_{\rm p})~{\rm d}^3{\bf x}\\
&=\frac{1}{8\pi}\int_{\Omega}(B_{\rm p}^2 + B_{\rm j}^2)~{\rm d}^3{\bf x}+ \frac{1}{4\pi}\int_{\Omega}{\bf B}_{\rm j} \cdot {\bf B}_{\rm p}~{\rm d}^3{\bf x}\\
&=\frac{1}{8\pi}\int_{\Omega}(B_{\rm p}^2 + B_{\rm j}^2)~{\rm d}^3{\bf x}+ \frac{1}{4\pi}\int_{\Omega}{\bf B}_{\rm j} \cdot \nabla\phi~{\rm d}^3{\bf x}\\
&=\frac{1}{8\pi}\int_{\Omega}(B_{\rm p}^2 + B_{\rm j}^2)~{\rm d}^3{\bf x}+ \frac{1}{4\pi}\int_{\Omega} \nabla\cdot(\phi{\bf B}_{\rm j})~{\rm d}^3{\bf x}\\
&=E_{\rm p}+E_{\rm j}+ \frac{1}{4\pi}\oint_{\Omega} \phi{\bf B}_{\rm j}\cdot\hat{\bf n}~{\rm d}{\rm S}.\\
\end{aligned}
\end{equation}
According to the decomposition (\ref{aeq:1}) and the boundary condition (\ref{aeq:2}), ${\bf B}_{\rm j}\cdot\hat{\bf n}$ is zero on all boundaries, thus the surface integral is zero.
Then the magnetic energy can be written as $E =E_{\rm j} + E_{\rm p}$, so that $E_{\rm p}$ is the minimum energy field meeting the boundary condition (\ref{aeq:2}).
This is the Thomson theorem for the above decomposition of the magnetic field.
This energy is achieved by reducing the current in the volume whilst preserving the normal component of the magnetic field on all boundaries.

Consider a periodic field ${\bf B}_{\rm periodic}$ in domain, $0\le x \le L_x$, $0\le y \le L_y$, and $0\le z \le L_z$.
The periodic boundary conditions are defined by
\begin{equation}\label{aeq:6}
\begin{aligned}
{\bf B}_{\rm periodic}\cdot\hat{\bf n}|_{x=0}={\bf B}_{\rm periodic}\cdot\hat{\bf n}|_{x=L_x}, \\
{\bf B}_{\rm periodic}\cdot\hat{\bf n}|_{y=0}={\bf B}_{\rm periodic}\cdot\hat{\bf n}|_{y=L_y}.
\end{aligned}
\end{equation}
We consider the the decomposition,
\begin{equation}\label{aeq:5}
{\bf B}_{\rm periodic} = {\bf B}_{\rm c} + {\bf B}_{\rm 0},
\end{equation}
where ${\bf B}_0$ is the potential field with Neumann boundary conditions on the top and bottom boundaries:
\begin{equation}\label{aeq:4}
{\bf B}_0\cdot\hat{\bf n}|_{z=0,~L_z}={\bf B}_{\rm periodic}\cdot\hat{\bf n}|_{z=0,~L_z},
\end{equation}
and periodic lateral boundary conditions:
\begin{equation}
\begin{aligned}
{\bf B}_0\cdot\hat{\bf n}|_{x=0}&={\bf B}_0\cdot\hat{\bf n}|_{x=L_x},\\
{\bf B}_0\cdot\hat{\bf n}|_{y=0}&={\bf B}_0\cdot\hat{\bf n}|_{y=L_y},
\end{aligned}
\end{equation}
and where ${\bf B}_{\rm c}$ is a non-potential field which is zero on the top and bottom boundaries:
\begin{equation}
{\bf B}_{\rm c}\cdot\hat{\bf n}|_{z=0,~L_z}=0,
\end{equation}
and which also has periodic lateral boundary conditions:
\begin{equation}
\begin{aligned}
{\bf B}_{\rm c}\cdot\hat{\bf n}|_{x=0}&={\bf B}_{\rm c}\cdot\hat{\bf n}|_{x=L_x},\\
{\bf B}_{\rm c}\cdot\hat{\bf n}|_{y=0}&={\bf B}_{\rm c}\cdot\hat{\bf n}|_{y=L_y}.
\end{aligned}
\end{equation}
The periodic potential field can also be written as the gradient of a scalar field ${\bf B}_0=\nabla\psi$.
Without affecting ${\bf B}_0$, we can neglect the the constant part of $\psi$, thus $\psi$ is a superposition of a linear function of $z$ and a sine and cosine function of $x$ and $y$ times an exponential function of $z$.
Then the solution of $\psi$ is also periodic in $x$ and $y$.
The details of the calculation of $\psi$ are given in Appendix \ref{sec:a:2}.

Considering the energy of this decomposition, similar to (\ref{aeq:3}), 
\begin{equation}
E=E_0+E_{\rm c}+ \frac{1}{4\pi}\oint_{\Omega} \psi{\bf B}_{\rm c}\cdot\hat{\bf n}~{\rm d}{\rm S}.\\
\end{equation}
The periodic potential field satisfies the condition \ref{aeq:4}, which lead to $\mathbf{B}_{\rm c}$ vanishing on the top and bottom boundaries, and the associated surface integral being zero.
As a result of this, the surface integral becomes
\begin{equation}\label{aeq:11}
\begin{aligned}
\oint_{\Omega} \psi{\bf B}_{\rm c}\cdot\hat{\bf n}~{\rm d}{\rm S}&=\int_{x=0} - B_{{\rm c},x}\psi~{{\rm d}y {\rm d}z} + \int_{x=L_x} B_{{\rm c},x}\psi ~{{\rm d}y {\rm d}z}\\
&+\int_{y=0} - B_{{\rm c},y}\psi~{{\rm d}x {\rm d}z} + \int_{y=L_y} B_{{\rm c},y}\psi~{{\rm d}x {\rm d}z},
\end{aligned}
\end{equation}
where $B_{{\rm c},x}$ and $B_{{\rm c},x}$ are the $x$ and $y$ components of the field ${\bf B}_{\rm c}$.
Since $\psi$ is periodic in $x$ and $y$ direction, the surface integral terms cancel, so the energy can be written as $E=E_{\rm c} + E_0$.
Hence $E_0$ is the minimum energy field subject to the boundary condition (\ref{aeq:4}).
This is the Thomson theorem for the decomposition (\ref{aeq:5}).
This energy is achieved by reducing the current from the volume whilst preserving the normal component of the magnetic field on the top and bottom boundaries, i.e.~$z=0$ and $z=L_z$, subject to the constraint of a periodic boundary condition in the $x$ and $y$ directions.

For the field ${\bf B}_{\rm periodic}$, both decompositions can be applied.
Then we can further make a separation of the potential field, ${\bf B}_{\rm p} = {\bf B}_{\rm p1} + {\bf B}_0$, where ${\bf B}_{\rm p1}$ is a potential field with zero normal value at the top and bottom boundaries, and a normal component matching ${\bf B}_{\rm c}$ on the lateral boundaries.
Therefore, ${\bf B}_{\rm p1}$ is also periodic in $x$ and $y$ directions, and hence the cross term of the energy between ${\bf B}_{\rm p1}$ and ${\bf B}_0$ will be zero, as shown by replacing ${\bf B}_{\rm c}$ with ${\bf B}_{\rm p1}$ in eq.~\ref{aeq:11}.
It follows that $E_{\rm p}=E_{\rm p1}+E_0$, where $E_{\rm p1}$ is the energy of the potential field ${\bf B}_{\rm p1}$.
Hence,
\begin{equation}
E_{\rm p} \ge E_0.
\end{equation}
Thus $E_0$ is a lower ``minimum" energy for a field matching ${\bf B}_{\rm periodic}$ on the bottom and top boundaries.
The demonstration of this lower ``minimum" energy state assumes that the original field $\bf B$ meets the lateral periodic boundary condition (\ref{aeq:6}).
However, in the numerical tests in Section \ref{sec:3}, using the TD model which is not periodic in the $x$ and $y$ directions, we also find that $E_0$ is smaller than $E_{\rm p}$ (Figure \ref{fig:2}(a) and Figure \ref{fig:4}(a)).
Also, in the test with half data of the magnetic field in Appendix \ref{sec:a:3}, the results show that $E_0$ is much smaller than $E_{\rm p}$ (Figure \ref{fig:7}(a)).
These results show that ${\bf B}_0$ can be a lower energy field than ${\bf B}_{\rm p}$ even when the total field is not periodic.

\section{Periodic solution of Laplace's Equation}
\label{sec:a:2}

To calculate the lateral periodic potential magnetic field in a rectangular domain ($0\leq x \leq L_x$, $0\leq y \leq L_y$, $0\leq z \leq L_z$) based on the bottom and top boundaries ($z=0,L_z$), we define the field as a gradient of a scalar function, ${\bf B}_0=\nabla \psi$.
Then $\psi$ satisfies the Laplace's equation $\nabla^2\psi=0$.
The method of separation of variables can be used, which means that we assume the solution is a superposition of all the basic separable solutions, $\psi=\sum_iX_i(x)Y_i(y)Z_i(z)$, where $X_i(x)$, $Y_i(y)$, and $Z_i(z)$ are functions which only depend on each coordinate, $i$ is the index of each basic separable solution.
The Laplace's equation becomes:
\begin{equation}
Y_i(y)Z_i(z)\frac{{\rm d^2}X_i(x)}{{\rm d}x^2} + X_i(x)Z_i(z)\frac{{\rm d^2}Y_i(y)}{{\rm d}y^2} + X_i(x)Y_i(y)\frac{{\rm d^2}Z_i(z)}{{\rm d}z^2} =0.
\end{equation}
This equation can be translated to three ordinary differential equations:
\begin{equation}
\begin{aligned}
\frac{{\rm d^2}X_i(x)}{{\rm d}x^2}&=-a^2X_i(x),\\
\frac{{\rm d^2}Y_i(y)}{{\rm d}y^2}&=-b^2Y_i(y),\\
\frac{{\rm d^2}Z_i(z)}{{\rm d}z^2}&=(a^2+b^2)Z_i(z).\\
\end{aligned}
\end{equation}
The periodic lateral boundary condition of the magnetic field restricts the values of $a$ and $b$ to the sets $\{\frac{2\pi n}{L_x}, n=0,1,2,3,...\}$ and $\{\frac{2\pi m}{L_y}, m=0,1,2,3,...\}$, respectively.
In the condition $n=m=0$, the solution sets for $X_i(x)$, $Y_i(y)$, and $Z_i(z)$ are:
\begin{equation}
\begin{aligned}
\{1, x\},
\{1, y\},
\{1, z\}.
\end{aligned}
\end{equation}
The solutions $xy$, $xz$, $yz$, and $xyz$ should be ruled out by the periodic magnetic field.
Moreover, the term $xy$ only contributes to an extra horizontal field, which cannot be determined from the Neumann boundary condition on the top and bottom boundaries.
Therefore, for simplicity and the above reason, we ignore the term $xy$.
If $n=0$ and $m\neq 0$, then the solution sets become:
\begin{equation}
\begin{aligned}
\{1\},
\{\cos(\frac{2\pi m}{L_y}y), \sin(\frac{2\pi m}{L_y}y)\},
\{\exp(-\frac{2\pi m z}{L_y}),\exp(-\frac{2\pi m z}{L_y}) \}.
\end{aligned}
\end{equation}
If $m=0$ and $n\neq 0$, then we have the solution sets:
\begin{equation}
\begin{aligned}
\{\cos(\frac{2\pi n}{L_x}x), \sin(\frac{2\pi n}{L_x}x)\},
\{1\},
\{\exp(-\frac{2\pi n z}{L_x}),\exp(-\frac{2\pi n z}{L_x}) \}.
\end{aligned}
\end{equation}
If $n\neq 0$ and $m\neq 0$, then the solution sets of $X(x)$, $Y(y)$, and $Z(z)$ are:
\begin{equation}
\begin{aligned}
\{\cos(\frac{2\pi n}{L_x}x), \sin(\frac{2\pi n}{L_x}x)\},
\{\cos(\frac{2\pi m}{L_y}y), \sin(\frac{2\pi m}{L_y}y)\},
\{\exp(-2\pi\eta_{m,n} z),\exp(2\pi\eta_{m,n} z)\},
\end{aligned}
\end{equation}
where $\eta_{m,n}=\sqrt{\frac{n^2}{L_x^2}+\frac{m^2}{L_y^2}}$.
It is worth noting that as the constraint from the lateral periodic boundary condition on the solution of the magnetic field, the solution requires the equal net fluxes on the top and bottom boundaries.

As the Laplace's equation is a linear equation, the solution of $\psi$ will be a superposition of a series of solutions as following:
\begin{equation}
\begin{aligned}
\psi(x,y,z)&=c_0+c_1x+c_2y+c_3z\\
&+\sum_{m=1}^{\infty} c_{4,m}\cos(\frac{2\pi m}{L_y}y)\exp(-\frac{2\pi m z}{L_y}) +\sum_{n=1}^{\infty} c_{4,n}\cos(\frac{2\pi n}{L_x}x)\exp(-\frac{2\pi n z}{L_x})\\
&+\sum_{m=1}^{\infty} c_{5,m}\sin(\frac{2\pi m}{L_y}y)\exp(-\frac{2\pi m z}{L_y}) +\sum_{n=1}^{\infty} c_{5,n}\sin(\frac{2\pi n}{L_x}x)\exp(-\frac{2\pi n z}{L_x})\\
&+\sum_{m=1}^{\infty} c_{6,m}\cos(\frac{2\pi m}{L_y}y)\exp(\frac{2\pi m z}{L_y}) +\sum_{n=1}^{\infty} c_{6,n}\cos(\frac{2\pi n}{L_x}x)\exp(\frac{2\pi n z}{L_x})\\
&+\sum_{m=1}^{\infty} c_{7,m}\sin(\frac{2\pi m}{L_y}y)\exp(\frac{2\pi m z}{L_y}) +\sum_{n=1}^{\infty} c_{7,n}\sin(\frac{2\pi n}{L_x}x)\exp(\frac{2\pi n z}{L_x})\\
&+\sum_{m=1,n=1}^{\infty}c_{8,m,n}\cos(\frac{2\pi n}{L_x}x)\cos(\frac{2\pi m}{L_y}y)\exp(-2\pi\eta_{m,n}z)\\
&+\sum_{m=1,n=1}^{\infty}c_{9,m,n}\cos(\frac{2\pi n}{L_x}x)\cos(\frac{2\pi m}{L_y}y)\exp(2\pi\eta_{m,n} z)\\
&+\sum_{m=1,n=1}^{\infty}c_{10,m,n}\cos(\frac{2\pi n}{L_x}x)\sin(\frac{2\pi m}{L_y}y)\exp(-2\pi\eta_{m,n} z)\\ 
&+\sum_{m=1,n=1}^{\infty}c_{11,m,n}\cos(\frac{2\pi n}{L_x}x)\sin(\frac{2\pi m}{L_y}y) \exp(2\pi\eta_{m,n} z)\\
&+\sum_{m=1,n=1}^{\infty}c_{12,m,n}\sin(\frac{2\pi n}{L_x}x)\cos(\frac{2\pi m}{L_y}y)\exp(-2\pi\eta_{m,n} z)\\
&+\sum_{m=1,n=1}^{\infty}c_{13,m,n}\sin(\frac{2\pi n}{L_x}x)\cos(\frac{2\pi m}{L_y}y)\exp(2\pi\eta_{m,n} z)\\
&+\sum_{m=1,n=1}^{\infty}c_{14,m,n}\sin(\frac{2\pi n}{L_x}x)\sin(\frac{2\pi m}{L_y}y)\exp(-2\pi\eta_{m,n} z)\\ 
&+\sum_{m=1,n=1}^{\infty}c_{15,m,n}\sin(\frac{2\pi n}{L_x}x)\sin(\frac{2\pi m}{L_y}y) \exp(2\pi\eta_{m,n} z),
\end{aligned}
\end{equation}
where the symbols \{$c_i$, $i=0,1,2,3$\}, \{$c_{i,m}$, $c_{i,n}$, $i=4,5,6,7$\}, and \{$c_{i,m,n}$, $i=8,9,10,11,12,13,14,15$\} indicate the superposition coefficients of each term, which should be determined by boundary condition.

We use the Neumann boundary condition on the top and bottom boundaries.
As $B_z=\partial_z\psi$, the data on the top and bottom boundaries only constrains the terms in the variable $z$.
It is obvious that the only terms that cannot be determined using $B_z|_{z=0,L_z}$ are $c_0$, $c_1x$, and $c_2y$.
The term $c_0$ makes no difference on the periodic potential field, while $c_1x$ and $c_2y$ give a constant horizontal field, which cannot be constrained by the Neumann boundary condition.
For simplicity, we set $c_1=c_2=0$, and hence the solution is unique if we ignore a constant horizontal field.
The other reason to ignore the constant horizontal field is that we aim to obtain a ``minimum'' energy potential field.
Finally, the term $c_3$ can be determined by the net flux on the bottom, $c_3=\frac{1}{L_xL_y}\int_{0}^{L_x}\int_{0}^{L_y}B_z(x,y){\rm d}x{\rm d}y$.

To determine the other coefficients, we define an inner product between two functions, $f_1$ and $f_2$, as $\langle f_1,f_2\rangle=\int_0^{L_x}\int_0^{L_y} f_1(x,y)f_2(x,y){\rm d}x{\rm d}y$.
It is obvious that the inner product between different terms is zero.
Thus the coefficients can be determined by taking the inner product between the bottom and top boundary normal component of the magnetic field with each term.
For example, the coefficients $c_{4,m}$ and $c_{6,m}$ can be determined by solving a linear equation:
\begin{equation}
\begin{aligned}
A_{11}c_{4,m} +A_{21}c_{6,m}&=\frac{2}{L_xL_y}\langle \cos(\frac{2\pi m}{L_y}y),B_{n,z=0}\rangle,\\
A_{12}c_{4,m} +A_{22}c_{6,m}&=\frac{2}{L_xL_y}\langle \cos(\frac{2\pi m}{L_y}y),B_{n,z=L_z}\rangle,
\end{aligned}
\end{equation}
where
\begin{equation}
\begin{aligned}
A_{11}&=-\frac{2\pi m}{L_y},\\
A_{21}&=\frac{2\pi m}{L_y},\\
A_{12}&=\frac{-2\pi m}{L_y}\exp(-2\pi m\frac{L_z}{L_y}),\\
A_{22}&=\frac{2\pi m}{L_y}\exp(2\pi m\frac{L_z}{L_y}).\\
\end{aligned}
\end{equation}
Here the terms $A_{11}$, $A_{12}$, $A_{21}$, and $A_{22}$ are corresponding terms of the matrix $A$, and the determinant of $A$ is $(\frac{2\pi m}{L_y})^2[\exp(-2\pi m\frac{L_z}{L_y}) - \exp(2\pi m\frac{L_z}{L_y})]$, which is not zero.
Hence the solution for $c_{4,m}$ and $c_{6,m}$ is uniquely determined.

For the coefficients $c_{8,m,n}$ and $c_{9,m,n}$, we can solve the linear equation:
\begin{equation}
\begin{aligned}
A_{11}c_{8,m,n} +A_{21}c_{9,m,n}&=\frac{4}{L_xL_y}\langle \cos(\frac{2\pi n}{L_x}x)\cos(\frac{2\pi m}{L_y}y),B_{n,z=0}\rangle,\\
A_{12}c_{8,m,n} +A_{22}c_{9,m,n}&=\frac{4}{L_xL_y}\langle \cos(\frac{2\pi n}{L_x}x)\cos(\frac{2\pi m}{L_y}y),B_{n,z=L_z}\rangle,
\end{aligned}
\end{equation}
where
\begin{equation}
\begin{aligned}
A_{11}&=-2\pi\eta_{m,n},\\
A_{21}&=2\pi\eta_{m,n},\\
A_{12}&=-2\pi\eta_{m,n}\exp(-2\pi\eta_{m,n} L_z),\\
A_{22}&=2\pi\eta_{m,n}\exp(2\pi\eta_{m,n} L_z).\\
\end{aligned}
\end{equation}
In this case the determinant of the matrix $A$ is $4\pi^2\eta_{m,n}^2[\exp(-2\pi\eta_{m,n} L_z) - \exp(2\pi\eta_{m,n} L_z)]$, which is not zero, which indicates that the coefficients can be determined uniquely.
The other coefficients can be determined similarly with the above calculation.

In summary, we have uniquely determined all the coefficients for $\psi$ according to the Neumann boundary condition on the top and bottom boundaries, and the periodic lateral boundary condition.
Therefore, the periodic potential field ${\bf B}_0=\nabla\psi$ can be uniquely calculated and only depends on $B_z|_{z=0,L_z}$.

\section{Method for computing $\mathbf A_{\rm \texorpdfstring{\MakeLowercase{P1}}{p1}}$ }
\label{sec_sag_appendix}

In this appendix we describe our method for computing
$\mathbf A_{\rm p1}$ in the Coulomb gauge for a current-free magnetic 
field in a box given the normal component of the magnetic field over 
the six planar boundaries of the box. 

\subsection{Domain and boundary-value problem for $\mathbf A_{\rm p1}$}
\label{sec_sag_domain}

We define a boundary-value problem for $\mathbf A_{\rm p1}$
in a Cartesian box with the normal component of $\mathbf B_{\rm c}$
prescribed on the boundary. Let $\Omega$ be the Cartesian box
\begin{equation}
  \Omega = \{ (x,y,z) | 0\le x \le L_x,
                             0\le y \le L_y,
                             0\le z \le L_z \},
\end{equation}
with boundary
\begin{equation}
  \partial\Omega = \bigcup S_{i},
\end{equation}
where $S_{i}$ are the six planar faces of the box. We 
label the faces by setting $i$ to a letter paired with a number,
e.g. $i = z1$. The letter is either $x$, $y$, or $z$ and 
indicates the normal direction to the boundary. The number
is either zero or one and indicates whether the boundary is the 
``lower'' or ``upper'' boundary respectively. For example, $S_{x 1}$ 
is the boundary at $x = L_x$, and $S_{z0}$ is the boundary at $z=0$.
 
In the interior of $\Omega$, let $\mathbf A_{\rm p1}$ satisfy the 
Coulomb gauge
\begin{equation}
  \nabla\cdot\mathbf A_{\rm p1} = 0.
  \label{equ_coulomb}
\end{equation}
In this case, a current-free ($\mathbf J=0$) 
magnetic field satisfies the vector Laplace's equation
\begin{equation}
  \nabla^2 \mathbf A_{\rm p1} = 0.
  \label{equ_vec_poisson}
\end{equation}
On the boundary $\partial\Omega$, we impose the boundary condition
\begin{equation}
  (\nabla\times\mathbf A_{\rm p1})\cdot \mathbf{\hat n} = \left. \mathbf B_{\rm c} \cdot 
                                                 \mathbf {\hat n} 
                                                 \right |_{\partial\Omega}.
\label{equ_bc_A}
\end{equation}
Equations (\ref{equ_vec_poisson}) - (\ref{equ_bc_A}) define the 
boundary-value problem for $\mathbf A_{\rm p1}$. 

The standard approach to formulating a boundary-value problem 
for the Laplace's equation is in terms of either Dirichlet or Neumann
boundary conditions \citep{morse1953methods}. In the context of 
computing $\mathbf A_{\rm p1}$, imposing
Dirichlet conditions corresponds to imposing
the transverse component of the vector potential:
\begin{equation}
  \mathbf A_t = \left. 
               (\mathbf A_{\rm p1} - 
               (\mathbf A_{\rm p1}\cdot\mathbf{\hat n}){\bf{\hat{n}}}) \right |_{ \partial\Omega},
\end{equation}
and imposing Neumann conditions corresponds to
specifying the normal derivative of the normal component:
\begin{equation}
  \partial_n A_n = \left.
                 \nabla (\mathbf A_{\rm p1} \cdot \mathbf{\hat n}) \cdot \mathbf{\hat n}
                 \right |_{\partial\Omega}.
\end{equation}
Equation (\ref{equ_bc_A}) does not directly match either 
of these forms, and hence it is necessary to derive a set of Dirichlet/Neumann boundary conditions
by first introducing additional gauge conditions at the boundary 
and secondly by solving a set of two-dimensional boundary-value
problems at each boundary $S_i$. By this means, a set of 
boundary data for $\mathbf A_t$ and $\partial A_n$ are derived
that are consistent with Equation (\ref{equ_bc_A}).
We describe this process in Section \ref{sec_res_bvp}. 

\subsection{Dirichlet/Neumann boundary conditions for $\mathbf A_{\rm p1}$
            for restricted distributions of $\mathbf B_{\rm c}$}
\label{sec_res_bvp}

Here we introduce additional gauge conditions to 
put the boundary-value problem described in 
Section \ref{sec_sag_domain} into
a standard Dirichlet/Neumann form. Our approach, however,
is ``restricted'' because it is only applicable when
$\mathbf B_{\rm c}$ satisfies the compatibility condition
\begin{equation}
  \int_{S_{i}} \mathbf B_{\rm c} \cdot \mathbf {\hat n} ~{\rm d}S = 0,
  \label{equ_compat}
\end{equation}
for all six boundary faces $S_{ij}$, i.e.~the net magnetic flux
over each individual face must be zero. This is a much more 
restrictive condition than the requirement of net flux
balance over the entire boundary, which should always be the 
case when $\nabla\cdot\mathbf B_{\rm c} = 0$. This restriction turns
out not to be a serious impediment however, as in Section
\ref{sec_sag_decom} we describe how the restricted approach
can be made applicable to a generic magnetic field through
the appropriate decomposition.

In addition to the Coulomb gauge condition, we follow 
\citet{1999A&A...350.1051A} and impose the further condition
\begin{equation}
  \left. \nabla_{i} \cdot \mathbf A_{\rm p1} \right |_{\partial\Omega} = 0.
  \label{equ_At_gcond}
\end{equation}
Here the operator $\nabla_{i} \cdot$ is a two-dimensional divergence operator 
defined on each face $i$. Given this constraint, it follows from 
Equations (\ref{equ_coulomb})-(\ref{equ_bc_A}) that
\begin{equation}
  \partial_n A_n = 0
  \label{equ_An_bc}
\end{equation} 
and
\begin{equation}
  \mathbf A_t = \nabla_{i} \chi_{i} \times \mathbf{\hat{n}},
  \label{equ_At_bc}
\end{equation}
where 
\begin{equation}
  \nabla^2_{i} \chi_{i} = \left. 
                          \mathbf B_{\rm c} \cdot \mathbf{\hat {n}} \right
                          |_{S_{i}}.
  \label{equ_chi_bn}
\end{equation}
Here again, the subscript $i$ indicates that the operator and
variable is defined on the two dimensional boundary plane $S_{i}$.

\noindent The boundary condition on $\mathbf A_t$ is computed by
solving Equation (\ref{equ_chi_bn}) on each boundary subject
to boundary conditions on each edge. The correct boundary conditions
are the homogeneous Neumann boundary conditions
\begin{equation}
  \partial_n \chi   = 0.
  \label{equ_chi_bcs}
\end{equation}
Equations (\ref{equ_chi_bcs}) and 
(\ref{equ_chi_bn}) define the boundary value problem for $\chi$
on each face. Since the boundary conditions are homogeneous 
Neumann boundary conditions, the source term
in Equation (\ref{equ_chi_bn}) must satisfy a compatibility 
condition \citep{Briggs:2000:MT:357695}. This condition is expressed by 
Equation (\ref{equ_compat}).

\subsection{Decomposition and solution for a generic magnetic field}
\label{sec_sag_decom}

In this subsection we describe how to decompose a generic
magnetic field so that problem of solving for $\mathbf A$ reduces
to solving the restricted boundary-value problem described
in Section \ref{sec_res_bvp}.

In order to satisfy the Neumann compatibility condition for a 
generic magnetic field, we decompose $\mathbf A$ as
\begin{equation}
  \mathbf A = \mathbf A^{\rm b} + \mathbf A^{\rm ub},
\end{equation}
where both $\mathbf A^{\rm b}$ and $\mathbf A^{\rm ub}$ must 
satisfy Equations (\ref{equ_vec_poisson}) and (\ref{equ_coulomb}).
We define $\mathbf A^{\rm ub}$ such that
\begin{equation}
  \int_{S_i} (\nabla\times \mathbf A^{\rm ub})\cdot \mathbf {\hat n}~{\rm d}S = 
  \int_{S_i} \mathbf B_{\rm c}\cdot \mathbf{\hat n}~{\rm d}S. 
  \label{equ_ub_def}
\end{equation} 
This condition ensures that
\begin{equation}
  \int_{S_i} (\nabla \times \mathbf  A^{\rm b})  \cdot \mathbf{\hat{n}}~{\rm d}S =
  \int_{S_i}   \mathbf B^{\rm b} \cdot \mathbf{\hat{n}}~{\rm d}S = 0 
  \label{equ_bal_cond}
\end{equation} 
over each boundary face.

The vector potential $\mathbf A^{\rm ub}$ is not uniquely defined by
Equation (\ref{equ_ub_def}) and can be chosen with some freedom.
For convenience, we choose a version of $\mathbf A^{\rm ub}$ with
a simple closed form expression. Its components are
\begin{equation}
   A^{\rm ub}_x = \frac{-\Phi_{z0} L_z y + (\Phi_{z1}-\Phi_{z0}) yz}{V},
\end{equation}
\begin{equation}
  A^{\rm ub}_y =  -\frac{\Phi_{x0} L_x z}{V}
\end{equation}
and
\begin{equation}
   A^{\rm ub}_z = \frac{-\Phi_{y0} L_y x + (\Phi_{x1}-\Phi_{x0}) xy}{V},
\end{equation}
where $V = L_x L_y L_z$, and 
\begin{equation}
  \Phi_i = \int_{S_i} \mathbf B_{\rm c} \cdot \mathbf{\hat{n}}~{\rm d}S
\end{equation}
is the net flux over the boundary $S_i$. When defining the flux,
we use the same normal on both the ``lower'' and 
``upper'' boundaries, e.g. the positive unit vector $\mathbf{\hat z}$
is used on both the $z=0$ and $z=L_z$ surfaces. 

\noindent It is straightforward to show that $\mathbf A^{\rm ub}$ satisfies
Equations (\ref{equ_vec_poisson}), (\ref{equ_coulomb}), and 
(\ref{equ_At_gcond}). The magnetic field corresponding to 
$\mathbf A^{\rm ub} $ has components
\begin{equation}
  \mathbf B^{\rm ub}_x = \frac{(L_x-x)\Phi_{x0} + \Phi_{x1}x}{V},
\end{equation}
\begin{equation}
  \mathbf B^{\rm ub}_y = \frac{(L_y-y)\Phi_{y0} + \Phi_{y1}y}{V},
\end{equation}
and
\begin{equation}
  \mathbf B^{\rm ub}_z = \frac{(L_z-z)\Phi_{z0} + \Phi_{z1}z}{V}.
\end{equation}
The divergence of this magnetic field is 
\begin{equation}
  \nabla\cdot \mathbf B^{\rm ub} = \frac{  \Phi_{x1}-\Phi_{x0}
                                   + \Phi_{y1}-\Phi_{y0}
                                   + \Phi_{z1}-\Phi_{z0}}{V},
\end{equation} 
and it follows that $\nabla\cdot \mathbf B^{\rm ub} = 0$ when 
there is net flux balance over the entire boundary $\partial\Omega$,
which is a basic requirement for any magnetic field \citep{1998clel.book.....J}.

\noindent Given $\mathbf B^{\rm ub}$, we may define a corrected
magnetic normal component 
\begin{equation}
  \mathbf B^{\rm b}\cdot\mathbf{\hat n} = \mathbf B_{\rm c} \cdot\mathbf{\hat n}
              - \mathbf B^{\rm ub} \cdot \mathbf{\hat{n}}.
  \label{equ_bn_corr}
\end{equation}
The vector potential $\mathbf A^{\rm b}$ can then be found by
the method of Section \ref{sec_res_bvp} with $\mathbf B^{\rm b}\cdot\mathbf{\hat n}$ as the 
right-hand side of Equation \ref{equ_chi_bn}. By construction
of $\mathbf B^{\rm ub}$, the Neumann compatibility condition is satisfied
for $\mathbf B^{\rm b}\cdot\mathbf{\hat n}$.

\subsection{Summary of method for computing $\mathbf A_{\rm p1}$}

\noindent Here we summarize our method for computing $\mathbf A_{\rm p1}$.

\begin{enumerate}
  \item Compute $\mathbf A^{\rm ub}$ and $\mathbf B^{\rm ub}$ analytically
        from $\mathbf B_{\rm c} \cdot \mathbf{\hat n}$.
  \item Compute $\mathbf B^{\rm b}\cdot\mathbf{\hat n}$ on the $\partial\Omega$ 
        from Equation (\ref{equ_bn_corr}).
  \item Compute $\mathbf A_t^{\rm b}$ on each face by solving 
        Equation (\ref{equ_chi_bn}) on each face $S_i$.
  \item Compute $\mathbf A^{\rm b}$ by solving the vector Laplace's equation
        with boundary conditions given by Equation (\ref{equ_An_bc})-(\ref{equ_At_bc}).
  \item Compute the resultant field $\mathbf A = \mathbf A^{\rm ub} + \mathbf A^{\rm b}$.
\end{enumerate}

The problem of determining $\mathbf A_{\rm p1}$ in the Coulomb
gauge in the context of computing helicity has been addressed
in a number of other works, e.g. \citet{2011SoPh..272..243T};
\citet{2011SoPh..270..165R}; \citet{2013SoPh..283..369Y}.
It is of some interest to compare our approach
to these. These methods, and ours, are similar in that they are 
based on the same gauge choice of \citet{1999A&A...350.1051A} at the boundary. 
One major difference between the methods is the treatment of 
the boundary-value problem for $\chi$. \citet{2011SoPh..272..243T} solve a 
nonhomogenous boundary-value problem for $\chi$ with $\partial_n\chi$ 
chosen on the edges to account for flux imbalance across each face.
\citet{2011SoPh..270..165R} perform a decomposition of $\mathbf A$
similar to that described in Section \ref{sec_sag_decom}.
Their choice of $\mathbf  A^{\rm b}	$, however, differs from ours. 
Our approach is simpler in a sense, because we do not need solve 
an algebraic system of determine our $\mathbf A$.

\subsection{Numerical implementation}
\label{sec_sag3}

We compute $\mathbf A_{\rm p1}$ by solving the vector
Laplace's equation using a numerical finite-difference method.
The problem is discretized using a second-order centered differencing
scheme \citep{Press:2007:NRE:1403886}. Both the two-dimensional
boundary-value problem for $\chi_i$ and the three-dimensional
boundary-value problem for $\mathbf A^{\rm b}$ are solved using
the same approach. 

The finite-difference equations are solved
using a geometric multigrid method  
with Red-Black relaxation as the basic relaxation operator
\citep{Briggs:2000:MT:357695,Press:2007:NRE:1403886}.
Our code performs multigrid V-cycles
until the maximum difference between V cycles is below a given
threshold. The method is implemented in Fortran2003 \citep{Metcalf:2011:MFE:2090092}
and all variables are stored in double precision. The code 
is parallelised for shared memory parallel computers using OpenMP \citep{Chandra:2001:PPO:355074}.

To demonstrate the method, we apply it to as simple 
analytic test case and measure the scaling of the numerical 
truncation error as a function of resolution. For a test case,
we consider the vector potential with components
\begin{equation}
  A_x = -A_0\cos(kx)\sin(ky)\exp(-lz),
\end{equation}
\begin{equation}
  A_y = +A_0\sin(kx)\cos(ky)\exp(-lz),
\end{equation}
and
\begin{equation}
  A_z =0,
\end{equation}
where $l = \sqrt{2}k$, and $A_0$ is a free parameter that we 
set to unity. For $k=2\pi n$, where $n$ is an integer, 
this vector potential satisfies the Coulomb gauge and the additional
gauge conditions at the boundary.

To measure the numerical error, we compare our numerical solution to
the analytic one using the following metrics
\begin{equation}
  E_{\rm max}(\mathbf V_1,\mathbf V_2) = \mbox{max}(|\mathbf V_1 - \mathbf V_2|),
\end{equation}
and
\begin{equation}
  E_{\rm avg}(\mathbf V_1,\mathbf V_2) = \langle |\mathbf V_1 - \mathbf V_2| \rangle,
\end{equation}
where $||$ is the component-wise absolute value, $\mbox{max}()$ is
the component-wise maximum over the whole domain, and $\langle \rangle$ is the average
over the domain.

\begin{figure}
  \includegraphics[width=.95\textwidth]{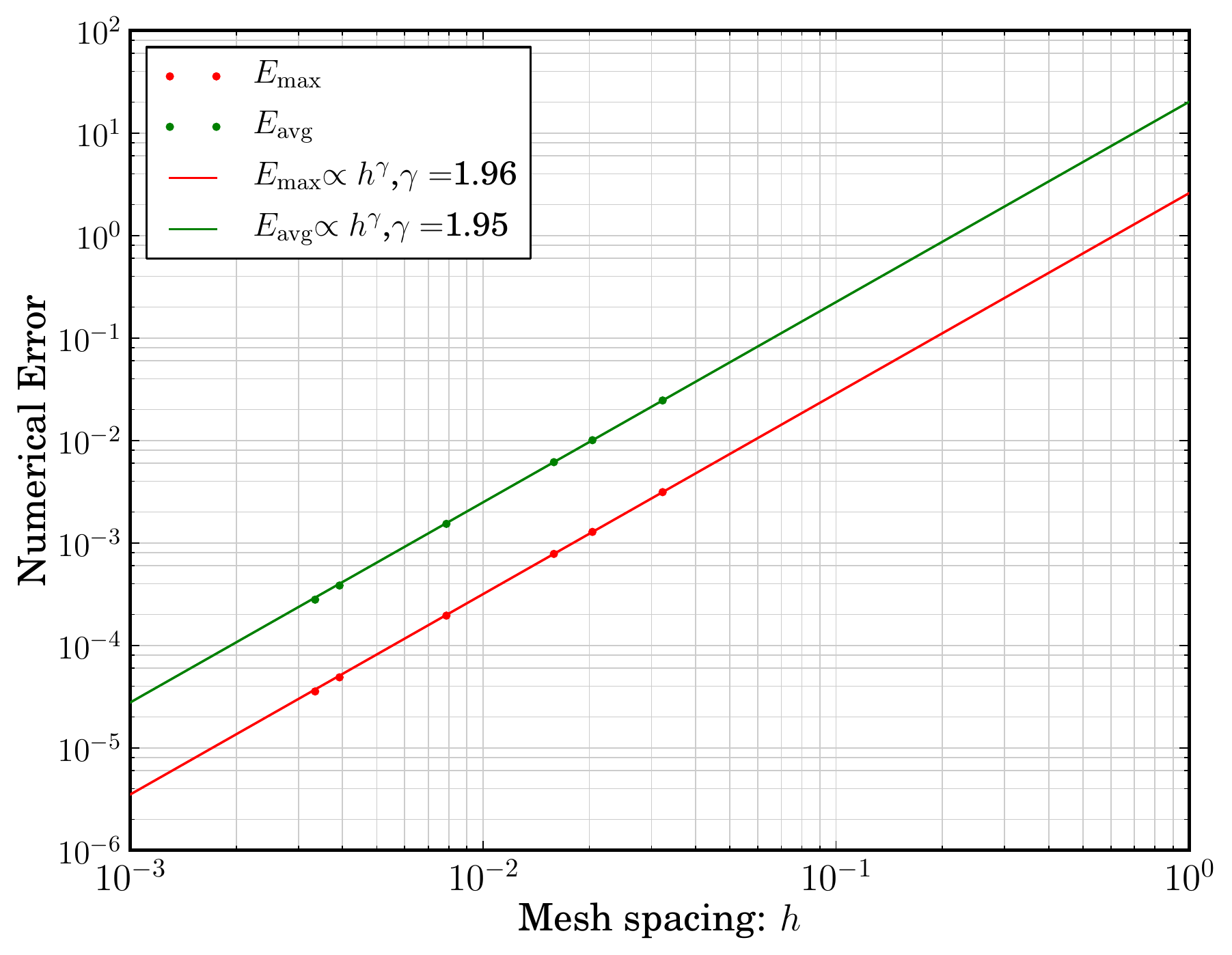}
  \caption{Numerical error versus mesh spacing $h$ for the 
           test case in Section \ref{sec_sag3}. The
           solid lines are power-law fits.}
  \label{fig_sag_num_error}
\end{figure}

Figure \ref{fig_sag_num_error} shows $E_{\rm max}$ and $E_{\rm avg}$ at different mesh 
spacings $h$ for a box of unity length in each direction. The 
solid lines are power-law fits to the data with power-law index
$\gamma$. Based on the fits, both metrics have scaling $\propto h^2$, 
which is consistent with the second-order discretization.

\section{Checking calculation by breaking the symmetries of the magnetic field}\label{sec:a:3}
\begin{figure}
\centering
\includegraphics[width=0.9\textwidth]{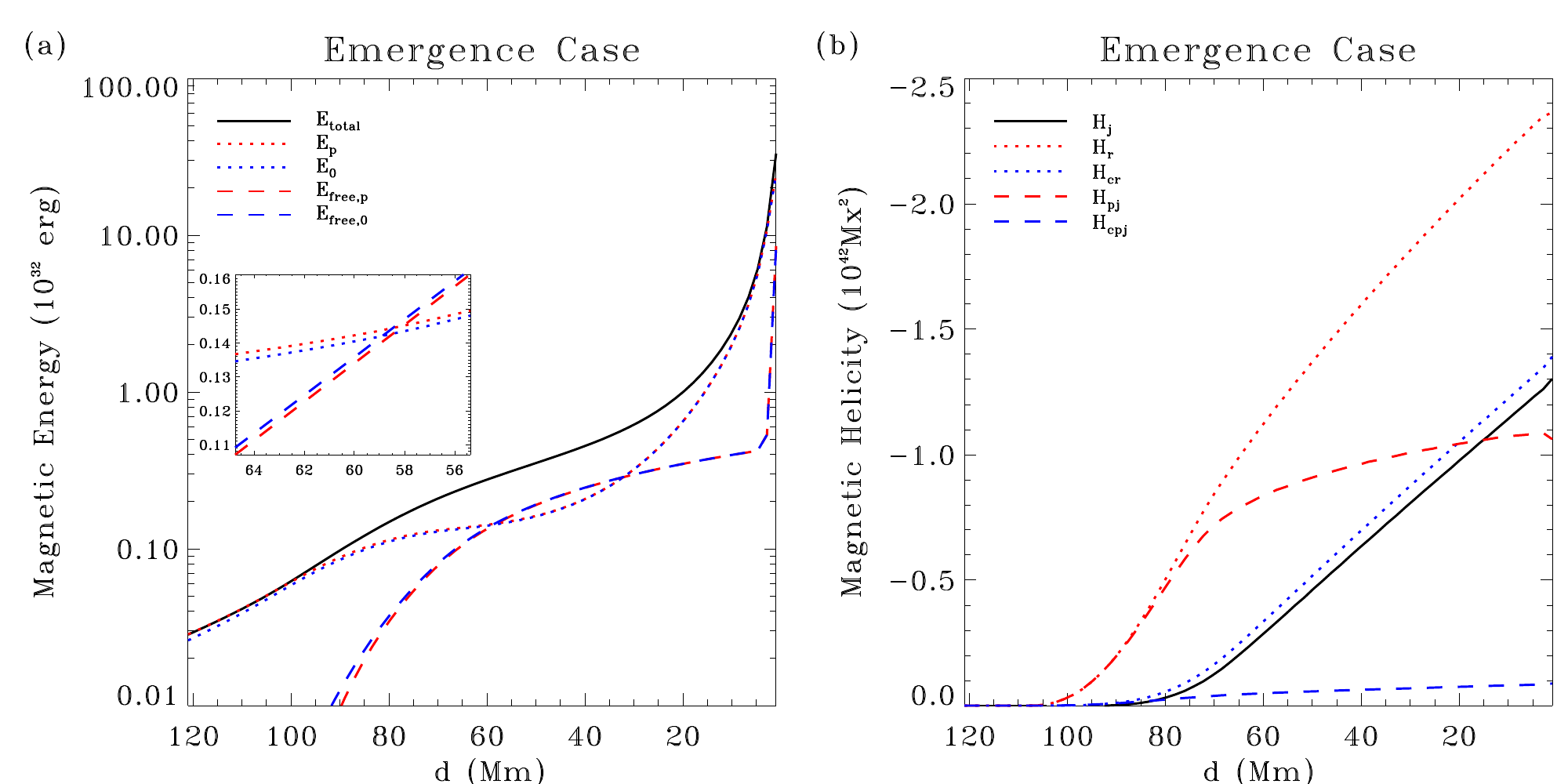}
\caption{(a) The black solid line indicates how the total magnetic energy evolves with the decrease of the parameter $d$ in the Titov-D\'{e}moulin sequence.
The red and blue dotted lines represent the energy of the potential field ($E_{\rm p}$) from the fixed boundary and that from the periodic boundary ($E_0$), respectively.
The red/blue dashed lines are the corresponding free energies, $E_{\text{free},\rm p}=E-E_{\rm p}$ and $E_{\text{free},0}=E-E_0$.
As the energies in the two cases are very close to each other, a sub-window shows the zoom-in view of a sub-range.
(b) The red dotted line is the usual relative magnetic helicity $H_{\rm r}$, and the blue dotted line is our new helicity $H_{\rm cr}$.
The components of $H_{\rm r}$ and $H_{\rm cr}$ are also shown: $H_{\rm j}$ (black solid), $H_{\rm pj}$ (red dash), and $H_{\rm cpj}$ (blue dash).}\label{fig:2}
\end{figure}

\begin{figure}
\centering
\includegraphics[width=0.9\textwidth]{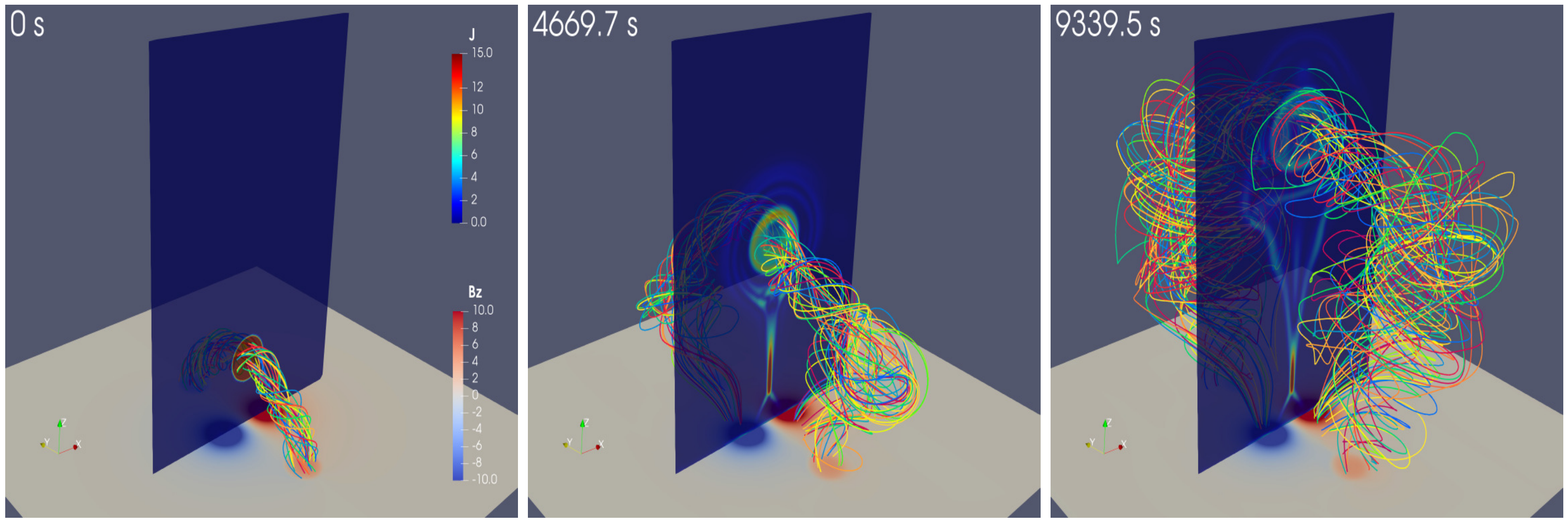}
\caption{The evolution of the magnetic field for the eruptive MHD case at three snapshots. 
The bottom boundary shows the distribution of $B_z$, the vertical slice shows the total current density $|{\bf J}|$, and the values of the magnetic field and current are displayed in the same way as in Figure \ref{fig:1}.
The colored lines indicate magnetic field lines associated with the erupting flux rope.}\label{fig:3}
\end{figure}

\begin{figure}
\centering
\includegraphics[width=0.9\textwidth]{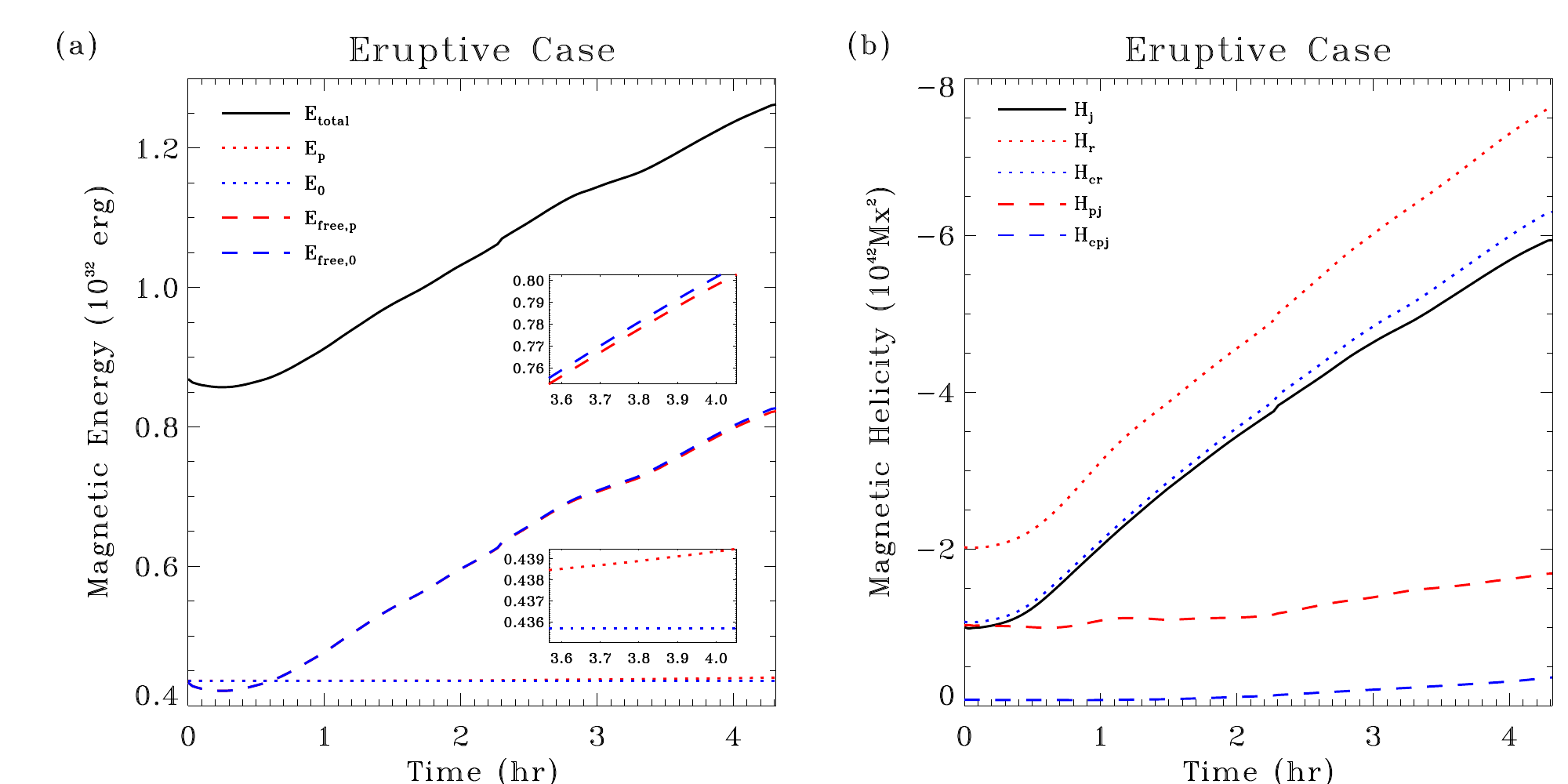}
\caption{(a) The black solid line indicates how the total magnetic energy evolves with time. 
The red and blue dotted lines represent the energy of the potential field ($E_{\rm p}$) from the fixed boundary and that from the periodic boundary ($E_0$), respectively.
The red/blue dashed lines are the corresponding free energies.
As the energies in the two cases are very close to each other, two sub-windows show a zoom-in view of a sub-range of the whole diagram.
(b) The red dotted line is the usual relative magnetic helicity $H_{\rm r}$, and the blue dotted line is our new helicity $H_{\rm cr}$.
The components of $H_{\rm r}$ and $H_{\rm cr}$ are also shown: $H_{\rm j}$ (black solid), $H_{\rm pj}$ (red dash), and $H_{\rm cpj}$ (blue dash).}\label{fig:4}
\end{figure}

\begin{figure}
\centering
\includegraphics[width=1\textwidth]{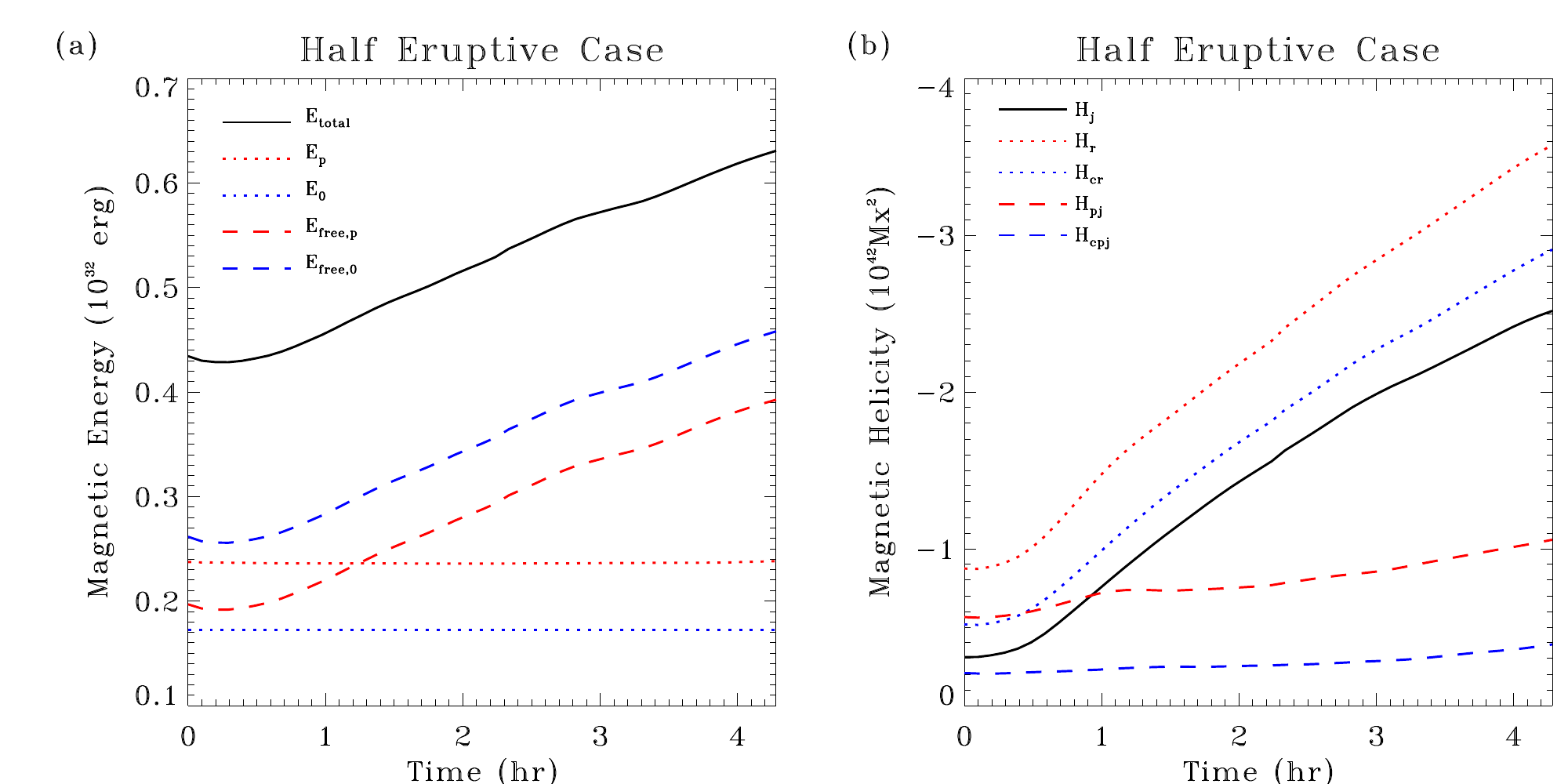}
\caption{The same as that in Figure \ref{fig:4} but the calculation is done in half of the original domain separated by a vertical plane (\textit{x--z}) at the middle of the computational box ($y=0$), which corresponds to the vertical plane cutting the flux rope shown in Figure \ref{fig:3}.}\label{fig:7}
\end{figure}

For the calculation in Section \ref{sec:3:1} and \ref{sec:3:2}, the computational domain is so large that the magnetic field on the side boundaries is very small.
In a more realistic case, the magnetic field on the side boundaries might not be small.
Moreover, the lower energy state of ${\bf B}_0$ mentioned in Section \ref{sec:a:1} might be not convincing enough given the small energy difference shown in Figure \ref{fig:2}(a) and Figure \ref{fig:4}(a).
Therefore, we apply the calculation on half of the original magnetic field by cutting the computational domain into two parts by a vertical plane (\textit{x--z}) at the middle of the computational box ($y=0$), which separates the flux rope into two equal parts and corresponds to the vertical plane shown in Figure \ref{fig:3}.
In this case the magnetic flux rope crosses the side boundary.
The evolution of the energy and helicity are shown in Figure \ref{fig:7}, from which we find that, as expected, the energy difference between the two potential fields $\mathbf{B}_{\rm p}$ and $\mathbf{B}_{0}$ is much larger than that shown in Section \ref{sec:3}.
This supports the lower energy state of ${\bf B}_0$ derived in Section \ref{sec:a:1}.
Nevertheless, the time evolution of each component of the magnetic energy and helicity still shows a similar behavior as that in Section \ref{sec:3}.

\newpage
\bibliographystyle{aasjournal}

\begin{thebibliography}{}
\expandafter\ifx\csname natexlab\endcsname\relax\def\natexlab#1{#1}\fi
\providecommand{\url}[1]{\href{#1}{#1}}
\providecommand{\dodoi}[1]{doi:~\href{http://doi.org/#1}{\nolinkurl{#1}}}
\providecommand{\doeprint}[1]{\href{http://ascl.net/#1}{\nolinkurl{http://ascl.net/#1}}}
\providecommand{\doarXiv}[1]{\href{https://arxiv.org/abs/#1}{\nolinkurl{https://arxiv.org/abs/#1}}}

\bibitem[{{Alissandrakis}(1981)}]{Alissandrakis1981}
{Alissandrakis}, C.~E. 1981, Astronomy and Astrophysics, 100, 197

\bibitem[{{Amari} {et~al.}(1999){Amari}, {Boulmezaoud}, \&
  {Mikic}}]{1999A&A...350.1051A}
{Amari}, T., {Boulmezaoud}, T.~Z., \& {Mikic}, Z. 1999, \aap, 350, 1051

\bibitem[{{Berger} \& {Rosner}(1995)}]{Berger1995}
{Berger}, M., \& {Rosner}, R. 1995, Geophysical and Astrophysical Fluid
  Dynamics, 81, 73, \dodoi{10.1080/03091929508229071}

\bibitem[{{Berger}(1984)}]{Berger1984b}
{Berger}, M.~A. 1984, Geophysical and Astrophysical Fluid Dynamics, 30, 79,
  \dodoi{10.1080/03091928408210078}

\bibitem[{{Berger}(1997)}]{Berger1997}
---. 1997, Journal of Geophysical Research, 102, 2637,
  \dodoi{10.1029/96JA01896}

\bibitem[{{Berger}(1999)}]{Berger1999}
---. 1999, Plasma Physics and Controlled Fusion, 41, B167,
  \dodoi{10.1088/0741-3335/41/12B/312}

\bibitem[{{Berger} \& {Field}(1984)}]{Berger1984a}
{Berger}, M.~A., \& {Field}, G.~B. 1984, Journal of Fluid Mechanics, 147, 133,
  \dodoi{10.1017/S0022112084002019}

\bibitem[{Briggs {et~al.}(2000)Briggs, Henson, \&
  McCormick}]{Briggs:2000:MT:357695}
Briggs, W.~L., Henson, V.~E., \& McCormick, S.~F. 2000, A Multigrid Tutorial
  (2Nd Ed.) (Philadelphia, PA, USA: Society for Industrial and Applied
  Mathematics), \dodoi{http://dx.doi.org/10.1137/1.9780898719505}

\bibitem[{Chandra {et~al.}(2001)Chandra, Dagum, Kohr, Maydan, McDonald, \&
  Menon}]{Chandra:2001:PPO:355074}
Chandra, R., Dagum, L., Kohr, D., {et~al.} 2001, Parallel Programming in OpenMP
  (San Francisco, CA, USA: Morgan Kaufmann Publishers Inc.)

\bibitem[{{Chiu} \& {Hilton}(1977)}]{Chiu1977}
{Chiu}, Y.~T., \& {Hilton}, H.~H. 1977, The Astrophysical Journal, 212, 873,
  \dodoi{10.1086/155111}

\bibitem[{{D{\'e}moulin} \& {Berger}(2003)}]{Demoulin2003}
{D{\'e}moulin}, P., \& {Berger}, M.~A. 2003, \solphys, 215, 203,
  \dodoi{10.1023/A:1025679813955}

\bibitem[{{Demoulin} {et~al.}(2006){Demoulin}, {Pariat}, \&
  {Berger}}]{Demoulin2006}
{Demoulin}, P., {Pariat}, E., \& {Berger}, M.~A. 2006, Solar Physics, 233, 3,
  \dodoi{10.1007/s11207-006-0010-z}

\bibitem[{Finn \& Antonsen(1985)}]{Finn1985}
Finn, J.~M., \& Antonsen, T. M.~J. 1985, 9, 111.
\newblock \url{http://inis.iaea.org/search/search.aspx?orig_q=RN:16069504}

\bibitem[{{Guo} {et~al.}(2017){Guo}, {Pariat}, {Valori}, {Anfinogentov},
  {Chen}, {Georgoulis}, {Liu}, {Moraitis}, {Thalmann}, \& {Yang}}]{Guo2017}
{Guo}, Y., {Pariat}, E., {Valori}, G., {et~al.} 2017, \apj, 840, 40,
  \dodoi{10.3847/1538-4357/aa6aa8}

\bibitem[{{Jackson}(1998)}]{1998clel.book.....J}
{Jackson}, J.~D. 1998, {Classical Electrodynamics, 3rd Edition}, 832

\bibitem[{{Jing} {et~al.}(2012){Jing}, {Park}, {Liu}, {Lee}, {Wiegelmann},
  {Xu}, {Deng}, \& {Wang}}]{Jing2012}
{Jing}, J., {Park}, S.-H., {Liu}, C., {et~al.} 2012, The Astrophysical Journal,
  752, L9, \dodoi{10.1088/2041-8205/752/1/L9}

\bibitem[{{Keppens} {et~al.}(2012){Keppens}, {Meliani}, {van Marle}, {Delmont},
  {Vlasis}, \& {van der Holst}}]{Keppens2012}
{Keppens}, R., {Meliani}, Z., {van Marle}, A.~J., {et~al.} 2012, Journal of
  Computational Physics, 231, 718, \dodoi{10.1016/j.jcp.2011.01.020}

\bibitem[{{Keppens} {et~al.}(2003){Keppens}, {Nool}, {T{\'o}th}, \&
  {Goedbloed}}]{Keppens2003}
{Keppens}, R., {Nool}, M., {T{\'o}th}, G., \& {Goedbloed}, J.~P. 2003, Computer
  Physics Communications, 153, 317, \dodoi{10.1016/S0010-4655(03)00139-5}

\bibitem[{{Linan} {et~al.}(2018){Linan}, {Pariat}, {Moraitis}, {Valori}, \&
  {Leake}}]{Linan2018}
{Linan}, L., {Pariat}, {\'E}., {Moraitis}, K., {Valori}, G., \& {Leake}, J.
  2018, The Astrophysical Journal, 865, 52, \dodoi{10.3847/1538-4357/aadae7}

\bibitem[{{Liu} {et~al.}(2014){Liu}, {Hoeksema}, {Bobra}, {Hayashi}, {Schuck},
  \& {Sun}}]{Liu2014}
{Liu}, Y., {Hoeksema}, J.~T., {Bobra}, M., {et~al.} 2014, \apj, 785, 13,
  \dodoi{10.1088/0004-637X/785/1/13}

\bibitem[{{Longcope} \& {Malanushenko}(2008)}]{2008Longcope}
{Longcope}, D.~W., \& {Malanushenko}, A. 2008, The Astrophysical Journal, 674,
  1130, \dodoi{10.1086/524011}

\bibitem[{{Low}(1996)}]{Low1996}
{Low}, B.~C. 1996, \solphys, 167, 217, \dodoi{10.1007/BF00146338}

\bibitem[{{Low}(2006)}]{Low2006}
---. 2006, \apj, 646, 1288, \dodoi{10.1086/504074}

\bibitem[{{Malanushenko} {et~al.}(2009){Malanushenko}, {Longcope}, {Fan}, \&
  {Gibson}}]{2009Malanushenko}
{Malanushenko}, A., {Longcope}, D.~W., {Fan}, Y., \& {Gibson}, S.~E. 2009,
  \apj, 702, 580, \dodoi{10.1088/0004-637X/702/1/580}

\bibitem[{{Mei} {et~al.}(2017){Mei}, {Keppens}, {Roussev}, \& {Lin}}]{Mei2017}
{Mei}, Z.~X., {Keppens}, R., {Roussev}, I.~I., \& {Lin}, J. 2017, \aap, 604,
  L7, \dodoi{10.1051/0004-6361/201731146}

\bibitem[{Metcalf {et~al.}(2011)Metcalf, Reid, \&
  Cohen}]{Metcalf:2011:MFE:2090092}
Metcalf, M., Reid, J., \& Cohen, M. 2011, Modern Fortran Explained, 4th edn.
  (New York, NY, USA: Oxford University Press, Inc.)

\bibitem[{{Moffatt}(1969)}]{Moffatt1969}
{Moffatt}, H.~K. 1969, Journal of Fluid Mechanics, 35, 117,
  \dodoi{10.1017/S0022112069000991}

\bibitem[{{Moraitis} {et~al.}(2019){Moraitis}, {Sun}, {Pariat}, \&
  {Linan}}]{Moraitis2019}
{Moraitis}, K., {Sun}, X., {Pariat}, {\'E}., \& {Linan}, L. 2019, \aap, 628,
  A50, \dodoi{10.1051/0004-6361/201935870}

\bibitem[{Morse \& Feshbach(1953)}]{morse1953methods}
Morse, P., \& Feshbach, H. 1953, Methods of theoretical physics, International
  series in pure and applied physics (McGraw-Hill).
\newblock \url{https://books.google.com/books?id=l8ENAQAAIAAJ}

\bibitem[{{Pariat} {et~al.}(2005){Pariat}, {D{\'e}moulin}, \&
  {Berger}}]{Pariat2005}
{Pariat}, E., {D{\'e}moulin}, P., \& {Berger}, M.~A. 2005, \aap, 439, 1191,
  \dodoi{10.1051/0004-6361:20052663}

\bibitem[{{Pariat} {et~al.}(2017){Pariat}, {Leake}, {Valori}, {Linton},
  {Zuccarello}, \& {Dalmasse}}]{Pariat2017}
{Pariat}, E., {Leake}, J.~E., {Valori}, G., {et~al.} 2017, \aap, 601, A125,
  \dodoi{10.1051/0004-6361/201630043}

\bibitem[{{Pariat} {et~al.}(2015){Pariat}, {Valori}, {D{\'e}moulin}, \&
  {Dalmasse}}]{Pariat2015}
{Pariat}, E., {Valori}, G., {D{\'e}moulin}, P., \& {Dalmasse}, K. 2015, \aap,
  580, A128, \dodoi{10.1051/0004-6361/201525811}

\bibitem[{Park {et~al.}(2010)Park, Chae, Jing, Tan, \& Wang}]{Park2010}
Park, S.-H., Chae, J., Jing, J., Tan, C., \& Wang, H. 2010, The Astrophysical
  Journal, 720, 1102, \dodoi{10.1088/0004-637x/720/2/1102}

\bibitem[{{Park} {et~al.}(2008){Park}, {Lee}, {Choe}, {Chae}, {Jeong}, {Yang},
  {Jing}, \& {Wang}}]{Park2008}
{Park}, S.-H., {Lee}, J., {Choe}, G.~S., {et~al.} 2008, \apj, 686, 1397,
  \dodoi{10.1086/591117}

\bibitem[{{Porth} {et~al.}(2014){Porth}, {Xia}, {Hendrix}, {Moschou}, \&
  {Keppens}}]{Porth2014}
{Porth}, O., {Xia}, C., {Hendrix}, T., {Moschou}, S.~P., \& {Keppens}, R. 2014,
  \apjs, 214, 4, \dodoi{10.1088/0067-0049/214/1/4}

\bibitem[{Press {et~al.}(2007)Press, Teukolsky, Vetterling, \&
  Flannery}]{Press:2007:NRE:1403886}
Press, W.~H., Teukolsky, S.~A., Vetterling, W.~T., \& Flannery, B.~P. 2007,
  Numerical Recipes 3rd Edition: The Art of Scientific Computing, 3rd edn. (New
  York, NY, USA: Cambridge University Press), 150

\bibitem[{{Prior} \& {MacTaggart}(2019)}]{2019Prior}
{Prior}, C., \& {MacTaggart}, D. 2019, Journal of Plasma Physics, 85,
  775850201, \dodoi{10.1017/S0022377819000229}

\bibitem[{{Prior} \& {Yeates}(2014)}]{2014Prior}
{Prior}, C., \& {Yeates}, A.~R. 2014, \apj, 787, 100,
  \dodoi{10.1088/0004-637X/787/2/100}

\bibitem[{{Rudenko} \& {Myshyakov}(2011)}]{2011SoPh..270..165R}
{Rudenko}, G.~V., \& {Myshyakov}, I.~I. 2011, \solphys, 270, 165,
  \dodoi{10.1007/s11207-011-9743-4}

\bibitem[{{Sakurai}(1979)}]{Sakurai1979}
{Sakurai}, T. 1979, \pasj, 31, 209

\bibitem[{{Stratton}(1941)}]{Stratton1941}
{Stratton}, J.~A. 1941, {Electromagnetic Theory} (John Wiley \& Sons, Inc.)

\bibitem[{{Taylor}(1986)}]{Taylor1986}
{Taylor}, J.~B. 1986, Reviews of Modern Physics, 58, 741,
  \dodoi{10.1103/RevModPhys.58.741}

\bibitem[{{Teunissen} \& {Keppens}(2019)}]{Teunissen2019}
{Teunissen}, J., \& {Keppens}, R. 2019, Computer Physics Communications, 245,
  106866, \dodoi{10.1016/j.cpc.2019.106866}

\bibitem[{{Thalmann} {et~al.}(2011){Thalmann}, {Inhester}, \&
  {Wiegelmann}}]{2011SoPh..272..243T}
{Thalmann}, J.~K., {Inhester}, B., \& {Wiegelmann}, T. 2011, \solphys, 272,
  243, \dodoi{10.1007/s11207-011-9826-2}

\bibitem[{{Titov} \& {D{\'e}moulin}(1999)}]{Titov1999}
{Titov}, V.~S., \& {D{\'e}moulin}, P. 1999, \aap, 351, 707

\bibitem[{{Valori} {et~al.}(2012){Valori}, {D{\'e}moulin}, \&
  {Pariat}}]{Valori2012}
{Valori}, G., {D{\'e}moulin}, P., \& {Pariat}, E. 2012, \solphys, 278, 347,
  \dodoi{10.1007/s11207-012-9951-6}

\bibitem[{{Valori} {et~al.}(2013){Valori}, {D{\'e}moulin}, {Pariat}, \&
  {Masson}}]{Valori2013}
{Valori}, G., {D{\'e}moulin}, P., {Pariat}, E., \& {Masson}, S. 2013, \aap,
  553, A38, \dodoi{10.1051/0004-6361/201220982}

\bibitem[{{Valori} {et~al.}(2016){Valori}, {Pariat}, {Anfinogentov}, {Chen},
  {Georgoulis}, {Guo}, {Liu}, {Moraitis}, {Thalmann}, \& {Yang}}]{Valori2016}
{Valori}, G., {Pariat}, E., {Anfinogentov}, S., {et~al.} 2016, Space Science
  Reviews, 201, 147, \dodoi{10.1007/s11214-016-0299-3}

\bibitem[{{Wheatland}(2006)}]{Wheatland2006}
{Wheatland}, M.~S. 2006, Solar Physics, 238, 29,
  \dodoi{10.1007/s11207-006-0232-0}

\bibitem[{{Wheatland}(2007)}]{Wheatland2007}
---. 2007, Solar Physics, 245, 251, \dodoi{10.1007/s11207-007-9054-y}

\bibitem[{{Woltjer}(1958{\natexlab{a}})}]{Woltjer1958a}
{Woltjer}, L. 1958{\natexlab{a}}, Proceedings of the National Academy of
  Science, 44, 489, \dodoi{10.1073/pnas.44.6.489}

\bibitem[{{Woltjer}(1958{\natexlab{b}})}]{Woltjer1958b}
---. 1958{\natexlab{b}}, Proceedings of the National Academy of Science, 44,
  833, \dodoi{10.1073/pnas.44.9.833}

\bibitem[{{Xia} {et~al.}(2018){Xia}, {Teunissen}, {El Mellah}, {Chan{\'e}}, \&
  {Keppens}}]{Xia2018}
{Xia}, C., {Teunissen}, J., {El Mellah}, I., {Chan{\'e}}, E., \& {Keppens}, R.
  2018, \apjs, 234, 30, \dodoi{10.3847/1538-4365/aaa6c8}

\bibitem[{{Yang} {et~al.}(2013{\natexlab{a}}){Yang}, {B{\"u}chner}, {Santos},
  \& {Zhang}}]{Yang2013}
{Yang}, S., {B{\"u}chner}, J., {Santos}, J.~C., \& {Zhang}, H.
  2013{\natexlab{a}}, Solar Physics, 283, 369,
  \dodoi{10.1007/s11207-013-0236-5}

\bibitem[{{Yang} {et~al.}(2013{\natexlab{b}}){Yang}, {B{\"u}chner}, {Santos},
  \& {Zhang}}]{2013SoPh..283..369Y}
---. 2013{\natexlab{b}}, \solphys, 283, 369, \dodoi{10.1007/s11207-013-0236-5}

\bibitem[{{Yang} {et~al.}(2018){Yang}, {B{\"u}chner}, {Sk{\'a}la}, \&
  {Zhang}}]{Yang2018}
{Yang}, S., {B{\"u}chner}, J., {Sk{\'a}la}, J., \& {Zhang}, H. 2018, Astronomy
  and Astrophysics, 613, A27, \dodoi{10.1051/0004-6361/201628108}

\bibitem[{{Zuccarello} {et~al.}(2018){Zuccarello}, {Pariat}, {Valori}, \&
  {Linan}}]{Zuccarello2018}
{Zuccarello}, F.~P., {Pariat}, E., {Valori}, G., \& {Linan}, L. 2018, \apj,
  863, 41, \dodoi{10.3847/1538-4357/aacdfc}

\end{thebibliography}

\newpage

\end{document}